\documentclass[prd,aps,preprint,groupedaddress,superscriptaddress,nofootinbib]{revtex4}
\usepackage{graphics}%
\usepackage[english]{babel}
\usepackage[mathcal]{euscript}
\usepackage[latin1]{inputenc}
\usepackage{latexsym}
\usepackage{amsmath}    
\usepackage{graphicx}   
\usepackage{verbatim}  
\usepackage{color}      
\usepackage{subfigure}  
\usepackage{bm}
\usepackage{amssymb}
\usepackage{hyperref}
\usepackage{bbm}
\usepackage{float}
\usepackage{amsfonts}
\usepackage{euscript}
\usepackage{amsmath}    
\usepackage{mathrsfs} 
\usepackage{bbding}
\usepackage{pifont}
\usepackage{cancel}
\usepackage{tikz}
 \usepackage[normalem]{ulem}
 
\usepackage{soul}

\begin{document}%

\title{Geodesically complete black holes in Lorentz-violating gravity}
\preprint{YITP-21-130}

\author{Ra\'ul Carballo-Rubio}
\affiliation{
Florida Space Institute, University of Central Florida, 
 12354 Research Parkway, Partnership 1, 32826 Orlando, FL, USA}
\author{Francesco Di Filippo}
\affiliation{Center for Gravitational Physics, Yukawa Institute for Theoretical Physics, Kyoto University, Kyoto 606-8502, Japan}
\affiliation{SISSA - International School for Advanced Studies, Via Bonomea 265, 34136 Trieste, Italy}
\author{Stefano Liberati}
\affiliation{SISSA - International School for Advanced Studies, Via Bonomea 265, 34136 Trieste, Italy}
\affiliation{
IFPU - Institute for Fundamental Physics of the Universe, Via Beirut 2, 34014 Trieste, Italy}
\affiliation{INFN Sezione di Trieste, Via Valerio 2, 34127 Trieste, Italy}
\author{Matt Visser}
\affiliation{
School of Mathematics and Statistics, Victoria University of Wellington, PO Box 600, Wellington 6140, New Zealand
}

\begin{abstract}
We present a systematic study of the geometric structure of non-singular spacetimes describing black holes in Lorentz-violating gravity. We start with a review of the definition of trapping horizons, and the associated notions of trapped and marginally trapped surfaces, and then study their significance in frameworks with modified dispersion relations. This leads us to introduce the notion of universally marginally trapped surfaces, as the direct generalization of marginally trapped surfaces for frameworks with infinite signal velocities (Ho\v rava-like frameworks), which then allows us to define universal trapping horizons. We find that trapped surfaces cannot be generalized in the same way, and discuss in detail why this does not prevent using universal trapping horizons to define black holes in Ho\v rava-like frameworks. We then explore the interplay between the kinematical part of Penrose's singularity theorem, which implies the existence of incomplete null geodesics in the presence of a focusing point, and the existence of multiple different metrics. This allows us to present a complete classification of all possible geometries that neither display incomplete physical trajectories nor curvature singularities. Our main result is that not all classes that exist in frameworks in which all signal velocities are realized in Ho\v rava-like frameworks. However, the taxonomy of geodesically complete black holes in Ho\v rava-like frameworks includes diverse scenarios such as evaporating regular black holes, regular black holes bouncing into regular white holes, and hidden wormholes.
\end{abstract}

\maketitle
\section{Introduction}

The recent dawn of gravitational wave astrophysics has lent further impetus to seeking a deeper understanding of black holes. In particular it has made pressing the necessity to provide predictions from  scenarios alternative to that offered by classical General Relativity (GR). While a large literature has been devoted to the study of black hole solutions in alternative theories of gravity, an interesting alternative is to consider black-hole like solutions where the inner singularity is regularized by quantum gravitational effects, as this is a general feature we do expect to be common to any quantum theory of gravity (at least in regards to singularities developing from a well posed Cauchy problem in classical GR, see e.g.~\cite{Horowitz:1995b}).   

This leads to the concept of regular (or non-singular) black holes \cite{Bardeen:1968,Dymnikova2004,Hayward2005,Ansoldi2008,Frolov2016,Konkowski:2018,Horowitz:1995a} or bouncing geometries \cite{Barcelo2014a, Barcelo2014b, Barcelo2015, DeLorenzo2015, Bianchi:2018}. In reference~\cite{Carballo-Rubio2019b} and the companion paper \cite{Carballo-Rubio2019a} (see also~\cite{Carballo-Rubio:2021ayp}), we showed that this class of metrics comprises all possible globally hyperbolic regular spacetimes describing the formation and disappearance of a black hole (technically, a trapped region), a result lending support to the use of these geometries in phenomenological studies. Noticeably, the classification in \cite{Carballo-Rubio2019a,Carballo-Rubio2019b} also describes alternative classes of metrics which however fail to satisfy some additional, reasonable, physical criteria, e.g. due to not being globally hyperbolic (or not even being everywhere analytical), or due to describing black holes incompatible with semiclassical physics (such as Hawking radiation).

An ansatz where the study of new features associated to a modified theory of gravity is naturally associated to a strong motivation for the study of regular and causally well posed geometries is that provided by Ho\v{r}ava--Lifshitz gravity \cite{Horava2009,Horava:2009b,Blas:2009a,Blas:2009b,Lu:2009,Charmousis:2009,Sotiriou:2009a,Sotiriou:2009b,Liberati:2013,Visser:2009,Jacobson:2010,Barausse:2011}. 
Indeed, this framework provides a specific example of a theory of gravity in which the violation of Lorentz invariance ensures at least power-counting renormalizablility~\cite{Visser:2009, Visser:power-count}: a feature which could lead to an improved short-distance behavior able to regularize spacetime singularities.\footnote{Recent work~\cite{Lara:2021jul} has illustrated that this is not the case at least for three spacetime dimensions and for a projectable version of Ho\v{r}ava--Lifshitz gravity. Also, it is possible that the presence of singularities in the geometries described in~\cite{Lara:2021jul} is an artifact of the spherically symmetric and stationary assumptions~\cite{Izumi:2009}.}
In this sense, Ho\v{r}ava--Lifshitz gravity or, more generally, Lorentz-violating theories with a preferred foliation, are well-motivated theoretical frameworks for extending our previously discussed general analysis of non-singular black hole geometries.  

Ho\v{r}ava--Lifshitz gravity is based on the introduction of an aether field which is constrained to be always timelike, of unit norm, and hypersurface orthogonal. The latter property is tantamount to saying that the aether can be written as the gradient of a scalar, the chronon, which can be used as the preferred time of the solutions of the theory. The Lagrangian then can be split in three terms depending on the mass  dimension of the operators making them up. In particular, the complete theory has mass dimension two, four and six operators (in 3+1 dimensions this is all one needs to ensure power counting renormalizability). 

The low-energy limit of Ho\v{r}ava--Lifshitz gravity (i.e.~the quadratic part of the Lagrangian) leads to a theory that has been previously proposed and independently analyzed, namely Einstein--aether theory~\cite{Jacobson:2000xp} (with the extra condition of hypersurface orthogonality of the aether~\cite{Jacobson:2010mx})\footnote{Normally, Einstein--aether theory, seen as an independent theory, does not require hypersurface orthogonality of the aether. However, we shall require it here as we are interested in the low energy limit of Ho\v{r}ava--Lifshitz gravity, whose power counting renormalizability is a motivation for exploring regular black hole solutions.}. See also references~\cite{Blas:2009ck,Jacobson:2010mx,Germani:2009yt}. 
Einstein--aether theory is defined as the most general quadratic theory of a unit timelike vector field coupled to a metric. As said, the unit timelike vector field defines a preferred frame (which, with the extra requirement of hypersurface orthogonality, is indeed a preferred foliation) throughout spacetime, and all the degrees of freedom of the theory propagate with finite speeds with respect to this preferred frame (trivially, a quadratic action in this setting can at most lead to relativistic dispersion relations for the propagating modes characterized by different, finite, limit speeds, i.e.~ of the form $\omega^2=c^2_{i}\, p^2$ for each $i$-mode). Hence, one of the main conceptual differences between Ho\v{r}ava--Lifshitz and Einstein--aether theories is the presence of unbounded propagation speeds in the former due to the higher dimensional terms in the action.

Interestingly, it is possible to obtain (in these theories with a preferred-foliation) solutions of the lowest-order action which describe black holes \cite{Blas2011,Barausse:2011pu}, i.e.~regions of spacetime which are causally disconnected in the future from the rest of the spacetime. This is surprisingly true even in the presence of unbounded propagation speeds, and due to the presence of a new structure inside the usual Killing horizon: the universal horizon. The latter can be readily understood as a compact constant-chronon surface, i.e.~a compact surface of simultaneity over which infinite speed signal are bound to propagate.  The fact that not even infinite speed signal can escape from within this boundary somewhat unexpectedly assures the survival of the notions of black hole and of (weak) cosmic censorship even in Ho\v{r}ava--Lifshitz gravity. However, very little is understood/known concerning regular black hole solutions in either this or the Einstein--aether setting. 

With this landscape in mind, in this paper we shall analyze the quasi-local structure of geodesically complete black holes in the above mentioned frameworks (preferred foliation with or without infinitely fast signals). We start in Sec.~\ref{sec:dsprel} with a discussion of signal velocities and the geometrical notions which are required to define black holes in frameworks in which these velocities are finite or infinite. In Sec.~\ref{sec:reg} we discuss the implications that the existence of focusing points have for theories with subluminal and superluminal modes. This forms the basis of the discussion of the possible classes of spacetimes in Sec.~\ref{sec:class}. Finally, we shall draw our conclusions and offer some future perspectives in Sec.~\ref{Concl}.
\section{Trapping horizons and modified dispersion relations \label{sec:dsprel}}

The study of the causal structure of modified gravity theories~\cite{Carballo-Rubio:2020ttr} shows that one must distinguish between frameworks in which all propagating modes have finite signal velocities and frameworks in which at least one propagating mode has an infinite signal velocity. Following the notation in reference~\cite{Carballo-Rubio:2020ttr}, we will refer to these two frameworks as Einstein--aether and Ho\v rava--like, respectively. We start by analyzing the former framework which, since all its signal velocities are finite, is closer to general relativity.

All spacetimes considered in this paper will be taken to be globally hyperbolic manifolds with orientable Cauchy surfaces. The spacetimes of interest describe the collapse of a regular distribution of matter from a given initial Cauchy surface with topology $\mathbb{R}^3$. On top of this structure we will define a metric $g_{ab}$ and a preferred vector field $n^a$, both assumed to satisfy suitable regularity conditions.

\subsection{Einstein--aether frameworks}

Let us start with the simplest setting containing both a metric $g_{ab}$ and a normalized aether field $n^a$, and construct the most general Lagrangian leading to 2nd-order field equations for both~\cite{Jacobson:2000xp,Jacobson:2008aj}:
\begin{equation}\label{eq:aeact}
\mathcal{L} = -\frac{c^4}{16\pi} R - K^{ab}{}_{mn} \;\nabla_a n^m \;\nabla_b n^n - \lambda(g_{ab} n^a n^b +1),
\end{equation}
where
\begin{equation}
K^{ab}{}_{mn} = c_1 g^{ab} g_{mn} 
+ c_2 \delta^a{}_m\delta^b_n + c_3 \delta^a{}_n \delta^b{}_m
+c_4 u^a u^b g_{mn}, 
\end{equation}
Note that we are using units in which $G=1$, but we are leaving $c$ explicit.

Focusing on the gravity-aether sector, Einstein--aether theory contains spin-2, spin-1, and spin-0 propagating modes. Different modes propagate with different signal velocities, which implies the existence of different horizons. These horizons can be defined, for instance using standard techniques in general relativity, but by working with the individual metrics
\begin{equation}\label{eq:metmod}
    g^{(i)}_{ab}=g_{ab}-\left(c_{(i)}^2/c^2-1\right)n_an_b,
\end{equation}
where $i$ can take the values 0, 1 and 2, and $\{c_{(i)}\}_{i=0}^2$ are the signal velocities of the propagating modes, while $c$ is the constant with dimensions of speed that appears in front of the Einstein-Hilbert part of the action defined in Eq.~\eqref{eq:aeact}. The discussion in this section is general enough so that we can consider arbitrary numbers of propagating modes with arbitrary, but finite, signal velocities.

\begin{figure}[!ht]%
\begin{center}
\vbox{\includegraphics[width=0.5\textwidth]{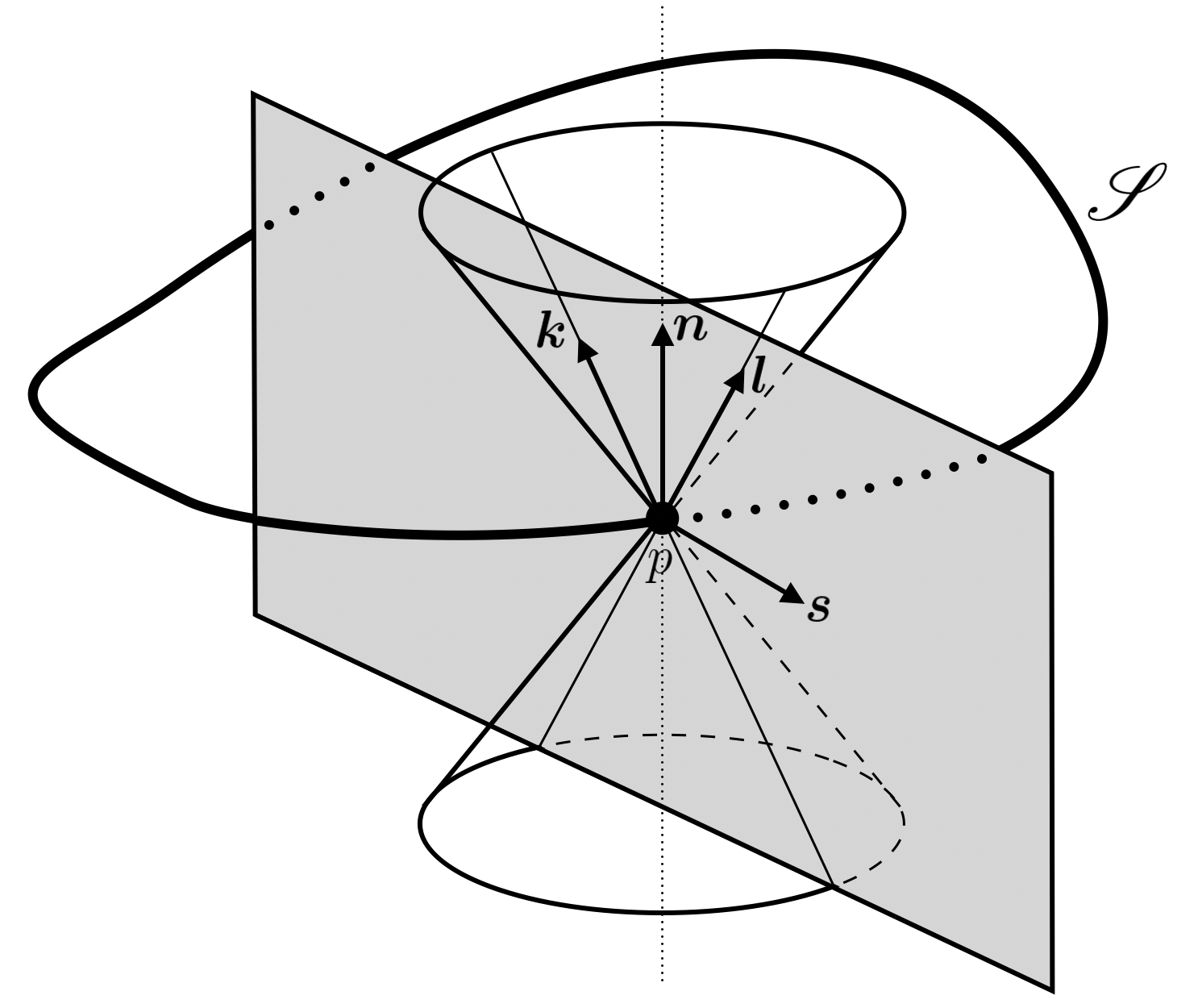}}
\bigskip%
\caption{Illustration of the closed 2-surface $\mathscr{S}$ and the null, timelike and spacelike directions orthogonal to it at a spacetime point $p$. For simplicity, $\mathscr{S}$ is represented as 1-dimensional (that is, we are removing one spacetime dimension). Adapted from~\cite{Gourgoulhon:2008pu}.}
\label{fig:vectors}
\end{center}
\end{figure}%

In this paper, we will always work with quasi-local characterizations of horizons and, in particular, with the related concepts of trapped surfaces~\cite{Penrose1964} and trapping horizons~\cite{Hayward1994}. Let us review some standard definitions in general relativity (see e.g.~\cite{Booth:2005qc,Gourgoulhon:2008pu} for reviews). Let us consider a closed 2-surface $\mathscr{S}$, spacelike for $g_{ab}$. The dimensionality of spacetime and the spacelike character of $\mathscr{S}$ implies the existence of two null directions of $g_{ab}$ that span the subspace of the tangent space orthogonal to this 2-surface (see Fig.~\ref{fig:vectors}). We can introduce two future directed null vectors $\bm{l}$ and $\bm{k}$ tangent to these null directions, following the usual convention that the former is outgoing and the latter ingoing. Assuming $\bm{l}\cdot \bm{k}=-1,$ the  2-metric orthogonal to the subspace spanned by the null vector fields $\{\bm{k},\bm{l}\}$ is given by
\begin{equation}\label{eq:htrans}
    h_{ab}=g_{ab}+k_a\,l_b+l_a\,k_b.
\end{equation}
Using the transverse metric, we can define the expansion scalars as
\begin{equation}\label{eq:nexpdef}
    \theta^{(\bm{X})}=\frac{1}{\sqrt{h}}\mathcal{L}_{\bm{X}}\sqrt{h},\qquad\qquad\bm{X}\in\{\bm{k},\bm{l}\}.
\end{equation}
These expansion scalars measure the fractional rate of change of the cross-sectional areas of the null congruences associated with $\bm{l}$ and $\bm{k}$~\cite{Poisson:2009pwt}.

The 2-surface $\mathscr{S}$ is called a \emph{future trapped surface} if
\begin{equation}\label{eq:tsdef}
    \left.\theta^{(\bm{k})}\right|_\mathscr{S}<0,\qquad \left.\theta^{(\bm{l})}\right|_\mathscr{S}<0,
\end{equation}
and a \emph{past trapped surface} if
\begin{equation}\label{eq:pastdef}
    \left.\theta^{(\bm{k})}\right|_\mathscr{S}>0,\qquad \left.\theta^{(\bm{l})}\right|_\mathscr{S}>0.
\end{equation}

There is an ambiguity in the definition of the vector fields $\{\bm{k},\bm{l}\}$, as the transformation $\bm{l}\rightarrow\alpha\bm{l}$, $\bm{k}\rightarrow\alpha^{-1}\bm{k}$ with $\alpha$ a positive function leaves $\bm{k}$ and $\bm{l}$ future directed null vectors, maintaining the normalization imposed above as well. However, the transverse metric in Eq.~\eqref{eq:htrans} is invariant under this transformation, while the expansion scalars change by the non-vanishing multiplicative factors 
$ \theta^{(\bm{l})} \to \alpha \theta^{(\bm{l})}$ and $ \theta^{(\bm{k})} \to \alpha^{-1} \theta^{(\bm{k})}$. In particular, the product $\theta^{(\bm{k})}\theta^{(\bm{l})}$ remains invariant. This means that the only relevant feature of each of these scalars is, when evaluated individually, their sign. This observation is compatible with the definitions in Eqs.~\eqref{eq:tsdef} and \eqref{eq:pastdef}.

The limiting case in which
\begin{equation}\label{eq:mts1}
    \left.\theta^{(\bm{k})}\right|_{\mathscr{S}}<0,\qquad \left.\theta^{(\bm{l})}\right|_{\mathscr{S}}=0.
\end{equation}
defines a \emph{future marginally trapped surface}. On the other hand, a \emph{past marginally trapped surface} is defined by
\begin{equation}\label{eq:pmts}
    \left.\theta^{(\bm{k})}\right|_{\mathscr{S}}>0,\qquad \left.\theta^{(\bm{l})}\right|_{\mathscr{S}}=0.
\end{equation}
In the following, we will focus on future trapped surfaces and, for notational simplicity, drop the adjective future when we refer to this and related definitions. However, one must keep in mind that the discussion below will have a mirrored one applying to past trapped surfaces. Although we will not replicate the complete discussion of past trapped surfaces below, we will introduce the most relevant definitions. We will always indicate explicitly when we are talking about past trapped surfaces to avoid confusion between the future and past cases.

Marginally trapped surfaces can be further classified depending on the value of the Lie derivative of $\theta^{(\bm{l})}$ along $\bm{k}$. A \emph{marginally outer trapped surface (MOTS)} satisfies the additional constraint
\begin{equation}\label{eq:singlemots}
    \left.\mathcal{L}_{\bm{k}}\theta^{(\bm{l})}\right|_{\mathscr{S}}<0,
\end{equation}
while a \emph{marginally inner trapped surface (MITS)} satisfies
\begin{equation}
    \left.\mathcal{L}_{\bm{k}}\theta^{(\bm{l})}\right|_{\mathscr{S}}>0.
\end{equation}
We can now consider a 3-surface foliated by MOTS, which following standard terminology (see \emph{e.g.} Refs.~\cite{Hayward1994,Andersson:2005gq}) we will call an \emph{outer trapping horizon (OTH)}. An outer trapping horizon provides a natural characterization of the boundary of a black hole that has found applications in diverse situations, as discussed for instance in~\cite{Andersson:2007fh} and references therein. If considering MITS, we would obtain an \emph{inner trapping horizon (ITH)} instead.

In this paper, we are dealing with Lorentz-violating theories in which the existence of a preferred frame is encapsulated in the existence of a preferred timelike vector field $\bm{n}$. Hence, it will be often useful to use the basis $\{\bm{n},\bm{s}\}$, where $\bm{n}$ is a future-pointing unit timelike vector and $\bm{s}$ is an outward pointing spacelike vector orthogonal to $\bm{n}$ (see Fig.~\ref{fig:vectors}). There always exist a choice of null vectors so that the two basis are related by the identities
\begin{equation}\label{eq:n-s}
    \bm{n}=\frac{\bm{l}+\bm{k}}{\sqrt{2}},\qquad \bm{s}=\frac{\bm{l}-\bm{k}}{\sqrt{2}}.
\end{equation}
Note that, once $\bm{n}$ is fixed, these relations fix the ambiguity discussed above in the definition of the null vectors. Also, the space-orientability of the spacetime manifold allow us to fix the sign of $\bm{s}$ so that it is outward-pointing (note that, due to the assumption of global hyperbolicity, we only need to do this in a specific Cauchy surface; this choice will then propagate consistently through time).

Using the identities relating the two bases, the transverse 2-metric $h_{ab}$ defined in Eq.~\eqref{eq:htrans} can be rewritten as
\begin{equation}\label{eq:transvmet}
    h_{ab}=g_{ab}+n_an_b-s_as_b\,.
\end{equation}
Hence, this is also the transverse metric to the subspace spanned by $\{\bm{n},\bm{s}\}$. By analogy with the corresponding definitions for the null vector fields $\{\bm{k},\bm{l}\}$, we can define the scalar quantities
\begin{equation}\label{eq:snexpdef}
    \theta^{(\bm{s})}=h^{ab}\nabla_a s_b=\frac{1}{\sqrt{h}}\mathcal{L}_{\bm{s}}\sqrt{h},\qquad\qquad\theta^{(\bm{n})}=h^{ab}\nabla_a n_b=\frac{1}{\sqrt{h}}\mathcal{L}_{\bm{n}}\sqrt{h}\, .
\end{equation}
These quantities are related to, but non-equivalent to, the standard definition of expansions for either spacelike or timelike congruences. This is because the latter are usually defined in terms of three-dimensional transverse metrics, which are different in the two cases. However, the quantities defined in Eq.~\eqref{eq:snexpdef} will be useful for the discussion below.

Let us now consider how the definitions above behave when considering a mode with finite signal velocity $c_{(i)}$. The null vector fields of the metric in Eq.~\eqref{eq:metmod} are now given by
\begin{equation}\label{eq:nullvcsc}
    \bm{l}^{(i)}=\frac{1}{\sqrt{2}}\left(\frac{c}{c_{(i)}}\bm{n}+\bm{s}\right),\qquad \bm{k}^{(i)}=\frac{1}{\sqrt{2}}\left(\frac{c}{c_{(i)}}\bm{n}-\bm{s}\right),
\end{equation}
or, equivalently, by
\begin{equation}\label{eq:inull}
    \bm{l}^{(i)}=\bm{l}-\frac{1}{\sqrt{2}}\left(1-\frac{c}{c_{(i)}}\right)\bm{n},\qquad \bm{k}^{(i)}=\bm{k}-\frac{1}{\sqrt{2}}\left(1-\frac{c}{c_{(i)}}\right)\bm{n}.
\end{equation}
These equations imply that
\begin{equation}\label{eq:sinv}
    \bm{n}^{(i)}=\frac{c}{c_{(i)}}\bm{n},\qquad \bm{s}^{(i)}=\bm{s}.
\end{equation}
On the other hand, the equivalent of Eq.~\eqref{eq:transvmet} becomes
\begin{equation}
 h^{(i)}_{ab}=g^{(i)}_{ab}+n^{(i)}_an^{(i)}_b-s^{(i)}_as^{(i)}_b,   
\end{equation}
which, taking into account that $n^{(i)}_a=c_{(i)}n_a/c$ and $s^{(i)}_a=s_a$, implies the equivalence
\begin{equation}\label{eq:transvmeteq}
 h^{(i)}_{ab}=h_{ab}.   
\end{equation}
With these equations we can deduce that
\begin{equation}\label{eq:nsexplk}
\theta^{\left(\bm{s}^{(i)}\right)}=\theta^{\left(\bm{s}\right)}=\frac{\theta^{\left(\bm{l}\right)}-\theta^{\left(\bm{k}\right)}}{\sqrt{2}},\qquad \theta^{\left(\bm{n}^{(i)}\right)}=\frac{c}{c_{(i)}}\theta^{\left(\bm{n}\right)}=\frac{c}{c_{(i)}}\left(\frac{\theta^{\left(\bm{l}\right)}+\theta^{\left(\bm{k}\right)}}{\sqrt{2}}\right),
\end{equation}
as well as
\begin{equation}\label{eq:likimts}
\theta^{\left(\bm{l}^{(i)}\right)}=\theta^{(\bm{l})}-\frac{1}{2}\left(1-\frac{c}{c_{(i)}}\right)\left[\theta^{(\bm{l})}+\theta^{(\bm{k})}\right],\qquad \theta^{\left(\bm{k}^{(i)}\right)}=\theta^{(\bm{k})}-\frac{1}{2}\left(1-\frac{c}{c_{(i)}}\right)\left[\theta^{(\bm{l})}+\theta^{(\bm{k})}\right].    
\end{equation}
It follows then that:
\begin{itemize}
    \item If $c_{(i)}>c$, a marginally trapped surface $\mathscr{S}$ of $g_{ab}$ is not a trapped surface of $g^{(i)}_{ab}$.
    \item If $c_{(i)}<c$, a marginally trapped surface $\mathscr{S}$ of $g_{ab}$ is also a trapped surface of $g^{(i)}_{ab}$.
    \end{itemize}
These statements are straightforward to prove using Eq.~\eqref{eq:likimts}. Indeed, for a marginally trapped surface of the metric $g_{ab}$, in which $\theta^{(\bm{l})}=0$, we have
\begin{equation}\label{eq:imetmts}
\theta^{\left(\bm{l}^{(i)}\right)}=-\frac{1}{2}\left(1-\frac{c}{c_{(i)}}\right)\theta^{(\bm{k})},\qquad \theta^{\left(\bm{k}^{(i)}\right)}=\frac{1}{2}\left(1+\frac{c}{c_{(i)}}\right)\theta^{(\bm{k})}.    
\end{equation}

Hence, the algorithm to define a black hole in this framework is as follows: select the mode with the highest signal velocity and consider its metric in Eq.~\eqref{eq:metmod} as well as its OTHs, which can be used to define the boundary of the black hole. For concreteness, let us denote this metric as $\hat{g}_{ab}$. In practice, it can be useful to perform a field redefinition of the metric and aether fields to new fields $\tilde{g}_{ab}$ and $\tilde{n}_a$ that makes $\hat{g}_{ab}=\tilde{g}_{ab}$, which always exists \cite{Barausse:2011pu}. The remaining metrics will also have OTHs, with larger radius the lower the associated signal velocity is, thus leading to a structure of nested OTHs that characterizes a black hole in Einstein-aether frameworks.

This procedure does not work in the presence of infinite signal velocities. In fact, from our discussion above it is clear that a marginally trapped surface of $g^{ab}$ (or any other metric associated with a mode with finite signal velocity) cannot be a marginally trapped surface of a mode with infinite signal velocity. More generally, a trapped surface of any metric $g^{(i)}_{ab}$ cannot be a marginally trapped surface of a mode with infinite signal velocity unless additional conditions are met. These additional conditions are studied in the next section.

\subsection{\protect{Ho\v rava}--like frameworks \label{sec:local}}

  For simplicity, let us consider a situation in which only one of the propagating modes has an infinite signal velocity. We will therefore have to study the limit $c_{(i)}/c\rightarrow \infty$, in which the expansions above are not defined, as the very metric in Eq.~\eqref{eq:metmod} ceases to be well-defined. However, this does not preclude the evaluation of their $c_{(i)}/c\rightarrow \infty$ limit. We will consider the most general case in which $c_{(i)}/c$ can take all possible values in $(0,\infty)$. In specific situations one expects to have a finite number of modes and therefore a discrete set of signal velocities, but this does not change the discussion and definitions below.

Our goal in this section is generalizing the definitions of trapped surface and marginally trapped surfaces to this situation. Let us start our exploration considering that $\mathscr{S}$ is a trapped surface for any metric $g^{(i)}_{ab}$ with arbitrary but finite $c_{(i)}$, namely
\begin{equation}\label{eq:mts}
    \left.\theta^{\left(\bm{k}^{(i)}\right)}\right|_{\mathscr{S}}<0,\qquad \left.\theta^{\left(\bm{l}^{(i)}\right)}\right|_{\mathscr{S}}<0,\qquad \forall c_{(i)}/c\in(0,\infty).
\end{equation}
These conditions imply certain constraints on the expansion scalars defined above. To see this explicitly, we can use Eq.~\eqref{eq:inull} which, combined with the linearity of the Lie derivative, then implies
\begin{equation}
    \theta^{\left(\bm{l}^{(i)}\right)}=\frac{1}{\sqrt{2}}\left(\frac{c}{c_{(i)}}\theta^{(\bm{n})}+\theta^{(\bm{s})}\right),\qquad \theta^{\left(\bm{k}^{(i)}\right)}=\frac{1}{\sqrt{2}}\left(\frac{c}{c_{(i)}}\theta^{(\bm{n})}-\theta^{(\bm{s})}\right).
\end{equation}
We can then write
\begin{equation}
    \theta^{\left(\bm{l}^{(i)}\right)}\theta^{\left(\bm{k}^{(i)}\right)}=\frac{1}{2}\left\{\left(\frac{c}{c_{(i)}}\right)^2\left[\theta^{(\bm{n})}\right]^2-\left[\theta^{(\bm{s})}\right]^2\right\}.
\end{equation}

That $\mathscr{S}$ is a trapped surface for all finite signal velocities and, in particular, $c_{(i)}=c$, implies that $\left.\theta^{(\bm{n})}\right|_{\mathscr{S}}<0$. This follows straightforwardly from the conditions that define a trapped surface in Eq.~\eqref{eq:mts}. The specific value of $\left.\theta^{(\bm{n})}\right|_{\mathscr{S}}$ is of geometric nature and does not depend on the value of the different signal velocities $c_{(i)}$, and we will assume that it is finite. Let us now consider different possible values of $\left.\theta^{(\bm{s})}\right|_{\mathscr{S}}$ and increase gradually the value of $c_{(i)}$ starting from $c_{(i)}=c$, analyzing the different behaviors that arise depending on the value of $\left.\theta^{(\bm{s})}\right|_{\mathscr{S}}$:

\begin{itemize}
    \item $\left.\theta^{(\bm{s})}\right|_{\mathscr{S}}\neq 0$: there exists a critical velocity $c_{(i)}=c_\star=-c\left.\theta^{(\bm{n})}\right|_{\mathscr{S}}/\left|\left.\theta^{(\bm{s})}\right|_{\mathscr{S}}\right|$ above which $\theta^{\left(\bm{l}^{(i)}\right)}\theta^{\left(\bm{k}^{(i)}\right)}<0$. Hence, modes with signal velocities higher than $c_\star$ are not trapped, while a mode with $c_{(i)}=c_\star$ is marginally trapped.
    \item $\left.\theta^{(\bm{s})}\right|_{\mathscr{S}}= 0$: all modes with finite signal velocities are trapped regardless of the value of $c_{(i)}$. Note that $c_\star/c\rightarrow\infty$ as $\left.\theta^{(\bm{s})}\right|_{\mathscr{S}}\rightarrow 0$.
\end{itemize}
Hence, $\mathscr{S}$ is a trapped surface for all propagating modes with finite (although unbounded) signal velocity if and only if $\left.\theta^{(\bm{s})}\right|_{\mathscr{S}}$ vanishes. With this information at hand we can introduce the following definitions:

\subsubsection{Universally marginally trapped surface:}

For a given metric $g^{(i)}_{ab}$, we can define a \emph{universally marginally trapped surface} as a closed 2-surface $\mathscr{S}$ satisfying
\begin{equation}\label{eq:umtsdef}
    \left.\theta^{\left(\bm{k}^{(i)}\right)}\right|_{\mathscr{S}}<0, \qquad     \left.\theta^{\left(\bm{l}^{(i)}\right)}\right|_{\mathscr{S}}<0, \qquad 
    \left.\theta^{(\bm{s})}\right|_{\mathscr{S}}=0.
\end{equation}
These conditions are equivalent to
\begin{equation}\label{eq:qluhdef2}
    \left.\theta^{\left(\bm{l}^{(i)}\right)}\right|_{\mathscr{S}}=\left.\theta^{\left(\bm{k}^{(i)}\right)}\right|_{\mathscr{S}}<0.
\end{equation}
With this definition we will recover the same quasi-local characterization of universal horizons proposed in Ref.~\cite{Maciel2015}. As we will discuss in more detail below, one can check that this definition reduces to the standard definition of a universal horizon in static situations~\cite{Maciel2015}.

First of all, let us remark that we can drop the index $(i)$ in the definition above. The reason is that both the spatial unit vector $\bm{s}$ and the expansion $\theta^{\left(\bm{s}^{(i)}\right)}$ are the same for all finite values of $c_{(i)}$, as made explicit in Eq.~\eqref{eq:nsexplk}. Equivalently, it is straightforward to check that, if Eq.~\eqref{eq:qluhdef2} is satisfied for one of the metrics, it will be then satisfied for all the remaining metrics $g^{(i)}_{ab}$. Hence, there is no need to specify the specific metric being considered when using this definition. 

An issue not mentioned in Ref.~\cite{Maciel2015} is that universally marginally trapped surfaces can be further classified as outer and inner, in a similar way as marginally trapped surfaces~\cite{Hayward1994}. The additional conditions can be deduced by imposing that $\mathscr{S}$ is surrounded by either marginally outer or inner trapped surfaces for propagating modes with arbitrarily high but finite velocity. In the former case, we have the following extension of Eq.~\eqref{eq:singlemots},
\begin{equation}
    \left.\mathcal{L}_{\bm{k}^{(i)}}\theta^{\left(\bm{l}^{(i)}\right)}\right|_{\mathscr{S}}<0,\qquad c/c_{(i)}=\epsilon\ll 1.
\end{equation}
Using Eq.~\eqref{eq:nullvcsc} and the linearity of the Lie derivative, we obtain 
\begin{equation}
   \epsilon^2\left.\mathcal{L}_{\bm{n}}\theta^{(\bm{n})}\right|_{\mathscr{S}}+\epsilon\left(\left.\mathcal{L}_{\bm{n}}\theta^{(\bm{s})}\right|_{\mathscr{S}}-\left.\mathcal{L}_{\bm{s}}\theta^{(\bm{n})}\right|_{\mathscr{S}}\right)-\left.\mathcal{L}_{\bm{s}}\theta^{(\bm{s})}\right|_{\mathscr{S}}<0.
\end{equation}
It follows that the definition of a \emph{universally marginally outer trapped surface (UMOTS)} entails, besides Eq.~\eqref{eq:umtsdef}, the additional condition
\begin{equation}\label{eq:umotsdef}
    \left.\mathcal{L}_{\bm{s}}\theta^{(\bm{s})}\right|_{\mathscr{S}}>0.
\end{equation}

Similarly, we define a \emph{universally marginally inner trapped surface (UMITS)} as satisfying the relation
\begin{equation}
    \left.\mathcal{L}_{\bm{s}}\theta^{(\bm{s})}\right|_{\mathscr{S}}<0.
\end{equation}
In full analogy with MOTS and MITS, we can define a \emph{universal outer trapping horizon (UOTH)} as a 3-surface foliated by UMOTS. Conversely, a \emph{universal inner trapping horizon (UITH)} is foliated by UMITS instead. We propose that the definition of UOTH provides an adequate quasi-local characterization of black holes in theories with infinite signal velocities (see Fig.~\ref{fig:nested}).

The existence of a UOTH/UITH is an invariant statement that does not depend on the specific metric $g^{(i)}_{ab}$ being considered. That is, if a UOTH/UITH is found for a metric $g^{(i)}_{ab}$, this 3-surface is also a UOTH/UITH for any other metric in Eq.~\eqref{eq:metmod}. This is reasonable from a conceptual perspective, given the universal (that is, applicable to all finite signal velocities) nature of this concept. On the other hand, the position of OTHs/ITHs change depending on the value of $c_{(i)}$ and, moreover, the fact that one exists for one value of $c_{(i)}$ does not imply the existence of a OTH/ITH for a different value of the signal velocity.

Only specific kinds of UOTHs have been (implicitly) discussed in the literature so far. The reason is that these appear naturally as the boundaries of black holes in Ho\v rava--like frameworks, in the same sense that OTHs do in general relativity. However, we also know from discussions in the framework of general relativity that OTHs are not enough in order to describe the internal structure of black holes, as ITHs are unavoidable if minimal regularity conditions are imposed \cite{Carballo-Rubio2019b}. We will find a similar situation in Lorentz-violating theories, as we will show that the description of non-singular black holes in these frameworks requires the existence of UITHs.

\subsubsection{Non-existence of universally trapped surfaces:}

While generalizing the notion of a marginally trapped surface to the notion of universally marginally trapped surface is quite natural as we have discussed above, generalizing the concept of a trapped surface is more subtle and, in fact, its straightforward generalization does not work. Naively generalizing Eq.~\eqref{eq:tsdef} to define a trapped region for the infinite velocity case as 
\begin{equation}\label{eq:utraplim}
    \lim_{c_{(i)}/c\rightarrow\infty}\theta^{\bm{l}^{(i)}}<0,\qquad \lim_{c_{(i)}/c\rightarrow\infty}\theta^{\bm{k}^{(i)}}<0,
\end{equation}
is inconsistent. In fact, a straightforward computation shows that
\begin{equation}\label{eq:utraplim2}
  \lim_{c_{(i)}/c\rightarrow\infty}\left(\theta^{\bm{l}^{(i)}}+\theta^{\bm{k}^{(i)}}\right)=0,
\end{equation}
which is incompatible with Eq.~\eqref{eq:utraplim}. The physical meaning of this incompatibility will be elaborated further below.

We can exploit the space-orientability of the manifold to define a \emph{trapped surface for the instantaneous mode} as a closed 2-surface $\mathscr{S}$ satisfying
\begin{equation}\label{eq:iumtsdef}
    \left.\theta^{\left(\bm{k}^{(i)}\right)}\right|_{\mathscr{S}}<0, \qquad     \left.\theta^{\left(\bm{l}^{(i)}\right)}\right|_{\mathscr{S}}<0, \qquad 
    \left.\theta^{(\bm{s})}\right|_{\mathscr{S}}<0.
\end{equation}
These conditions are equivalent to
\begin{equation}\label{eq:qluhdef3}
    \left.\theta^{\left(\bm{l}^{(i)}\right)}\right|_{\mathscr{S}}<\left.\theta^{\left(\bm{k}^{(i)}\right)}\right|_{\mathscr{S}}<0.
\end{equation}
The space-orientability of the manifold allows us to attach a meaning to the sign of $\left.\theta^{(\bm{s})}\right|_{\mathscr{S}}<0$, which is essential for the definition above. This definition implies that $\mathscr{S}$ is a trapped surface for the outward-pointing mode with infinite signal velocity, also known as instananeous mode. In fact, and as we will discuss in more detail in the specific examples below, this condition forces chronon hypersurfaces to have decreasing values of the areal radius in the outward direction.

\begin{figure}[!ht]%
\begin{center}
\vbox{\includegraphics[width=0.5\textwidth]{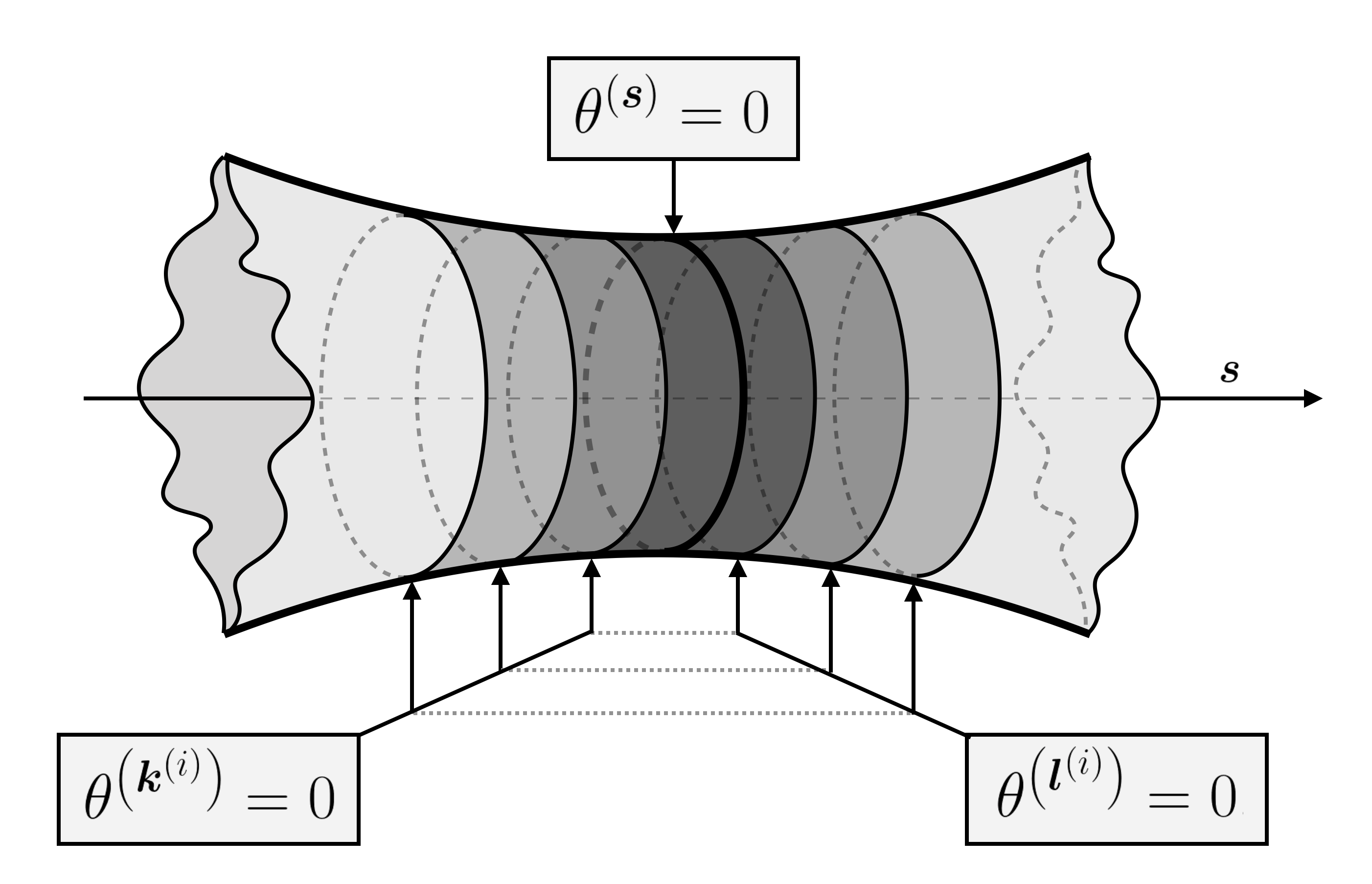}}
\bigskip%
\caption{Nested structure of black holes in Ho\v rava--like frameworks. This figure represents a portion of a constant-chronon hypersurface around an UMOTS, with the circular sections representing the areal radius $r$ and the outward direction indicated by $\bm{s}$. In theories with infinite signal velocities, what in general relativity is a thin boundary (a single MOTS) becomes a thick transition region that goes from the outermost MOTS to the UMOTS, in which $\theta^{(\bm{s})}=0$. Between the outermost MOTS and the UMOTS, there are formally infinite MOTS associated with the metrics $g^{(i)}_{ab}$, satisfying $\theta^{\left(\bm{l}^{(i)}\right)}=0$ with $c_{(i)}/c$ going from its lowest allowed value to infinite. Inside the UOTH we find a reverse nested structure but with $\theta^{\left(\bm{k}^{(i)}\right)}$ vanishing instead of $\theta^{\left(\bm{l}^{(i)}\right)}$, so that the whole structure is symmetric around the UMOTS. This implies that, inside the UMOTS, there are regions in which the areal radius can increase along the null direction $\bm{k}^{(i)}$ for large enough values of the signal velocity, which is possible due to the areal radius increasing towards the interior along a constant-chronon hypersurface (that is, in the direction of $-\bm{s}$), as well as the null direction $\bm{k}^{(i)}$ tending towards the inward-pointing spacelike vector $-\bm{s}$ as the value of $c_{(i)}/c$ increases.}
\label{fig:nested}
\end{center}
\end{figure}%

It is important to keep in mind that, while this definition can provide a useful characterization of the regions of spacetime bounded by universally marginally trapped surfaces, it is not equivalent to a ``universally trapped surface" in the following sense. This surface, while trapping the outward-pointing mode with infinite signal velocity (the instantaneous mode), in fact, all outward-pointing propagating modes, is not a trapped surface for all ingoing modes (see Fig.~\ref{fig:nested}). The first example is the inward-pointing mode with infinite signal velocity, for which the radius increases during a certain interval of time. Less trivial examples are propagating modes with finite but arbitrarily high signal velocities. From the equations above, for instance Eq.~\eqref{eq:utraplim2}, we can deduce that for high enough values of the signal velocity the expansion $\theta^{\left(\bm{k}^{(i)}\right)}$ always becomes positive. Hence, ingoing null rays are not trapped for high enough values of the signal velocity. This is reasonable as the trajectories of these modes approach the trajectory of the inward-pointing mode with infinite signal velocity as $c_{(i)}$ increases. This behavior appears only for ingoing modes; if we focus only on outgoing modes, the behavior is the one expected for a trapped surface and is, moreover, universal (that is, independent of the signal velocity).

In practice, this implies that there always exist high enough values of the signal velocity $c_{(i)}$ so that subsets of the region bounded by universally marginally trapped surfaces cannot be distinguished from the exterior region just by doing quasi-local experiments restricted to the interior region (as we have discussed above, is possible to break this degeneracy between interior and exterior by adding information about the orientation of the constant-chronon hypersurfaces). However, this region is still trapped in the looser sense that, being surrounded by universally marginally trapped surfaces, the areal radius is bounded from above for propagating modes in this region regardless of their signal velocity (for which the assumption of the topology of the constant-chronon hypersurfaces being $\mathbb{R}^3$ is key).\footnote{Something qualitatively similar takes place in geometries describing regular black holes, in which the core is not trapped but is nevertheless surrounded by trapped surfaces. Hence, while propagating modes inside the core can move in the direction of increasing areal radius during finite intervals of time, only a bounded value of the latter can be reached while the trapped surfaces last.} This aspect, that sets apart back holes in Ho\v rava-like frameworks, will be discussed in more detail below when considering specific examples.

\subsubsection{Regarding white holes:}

The discussion above has been focused on future trapped surfaces, but with a couple of sign changes it can be modified to describe past trapped surfaces, with the starting point being the definition in Eq.~\eqref{eq:pmts}. Following this procedure allow us to define a \emph{(past) universally marginally trapped surface} as a closed 2-surface $\mathscr{S}$ satisfying
\begin{equation}\label{eq:pumtsdef}
    \left.\theta^{(\bm{k})}\right|_{\mathscr{S}}>0, \qquad     \left.\theta^{(\bm{l})}\right|_{\mathscr{S}}>0, \qquad 
    \left.\theta^{(\bm{s})}\right|_{\mathscr{S}}=0.
\end{equation}
This concept provides a quasi-local characterization of white holes in Ho\v rava--like frameworks. This would also allow us to construct bouncing geometries that combine past and future universally marginally trapped surfaces. Note that, in this definition, we have already taking into account the invariance of $\theta^{(\bm{s})}$, and therefore we have not indicated explicitly the specific propagating mode and corresponding metric being considered.

\subsubsection{Specific coordinates}

For concreteness, let us particularize the discussion above for a specific system of coordinates. Given our assumptions, it is natural to use the preferred time function $\tau(x)$ as the temporal coordinate. On the other hand, spherical symmetry implies the existence of a second foliation defined by the area coordinate $r(x)$, with area of the spherically symmetric slices given by $A(x) = 4\pi r(x)^2$. This suggests using $r(x)$ as the second coordinate, which would imply using \{$\nabla_a\tau(x)$, $\nabla_a r(x)$\} as a basis for the radial-temporal cotangent space. A potential issue associated with this choice is that $\nabla_a\tau(x) \propto \nabla_a r(x)$ at the universal horizon, so the proposed basis fails there. 

A possible solution is to replace $r(x)$ with the proper distance to the centre of spherical symmetry, $\ell(x)$, to be  as measured along a specific leaf of the foliation $\Sigma_\tau$. In the $(\tau,\ell,\theta,\phi)$ coordinates, by construction $g_{\ell\ell}=1$ and $g^{\tau\tau} <1$, so that the complete metric reads
\begin{equation}\label{eq:metricc}
g_{ab} = \left[ \begin{array}{cc|cc} -N(\tau,\ell)^2 + v(\tau,\ell)^2 &\;\;v(\tau,\ell)\;\; &0&0\\ 
v(\tau,\ell) &1 & 0 &0 \\ \hline
0&0&r(\tau,\ell)^2 & 0\\0&0&0&r(\tau,\ell)^2\sin^2\theta\end{array}\right],
\end{equation}
where we have defined $g_{\tau\tau}=-N(\tau,\ell)^2 + v(\tau,\ell)^2$ and $g_{\tau\ell}=g_{\ell\tau}=v(\tau,\ell)$, so that
\begin{equation}\label{eq:metricc2}
g^{ab} = \left[ \begin{array}{cc|cc} -N(\tau,\ell)^{-2}&v(\tau,\ell)/N(\tau,\ell)^2 &0&0\\ 
v(\tau,\ell)/N(\tau,\ell)^2 &\;\;1-v(\tau,\ell)^2/N(\tau,\ell)^2\;\; & 0 &0 \\ \hline
0&0&r(\tau,\ell)^{-2} & 0\\0&0&0&r(\tau,\ell)^{-2}\sin^{-2}\theta\end{array}\right].
\end{equation}
Note $g_{\ell\ell}=1$ and $g^{\tau\tau} <1$ as required.

Hence, it will be often useful to use the basis $\{\bm{n},\bm{s}\}$, where $\bm{n}$ is the future-pointing unit timelike vector
\begin{equation}
    n^a=-\frac{1}{\sqrt{{-} g^{00}}}\;g^{ab}\; \nabla_b\tau,
\end{equation}
and $\bm{s}$ is {an outward pointing}  spacelike vector orthogonal to $\bm{n}$. In these coordinates, the vector fields $\{\bm{n},\bm{s}\}$ are given by
\begin{equation}
    \bm{n}=\frac{1}{N}\left(\partial_\tau-v\partial_\ell\right)
\end{equation}
and
\begin{equation}
    \bm{s}=\partial_\ell.
\end{equation}
Note that $N=N(\tau,\ell)$ has been defined implicitly in Eqs.~\eqref{eq:metricc} and \eqref{eq:metricc2}. The associated null vectors $\{\bm{k},\bm{l}\}$ are proportional to
\begin{equation}
    \bm{k}\propto \partial_\tau-(N+v)\partial_\ell
\end{equation}
and
\begin{equation}
    \bm{l}\propto \partial_\tau+(N-v)\partial_\ell.
\end{equation}
The condition $\bm{n}=(\bm{k}+\bm{l})/\sqrt{2}$ introduced in Eq.~\eqref{eq:n-s} fixes the normalization of the null vectors as
\begin{equation}
    \bm{k}=\frac{1}{\sqrt{2}N}\left[ \partial_\tau-(N+v)\partial_\ell\right]
\end{equation}
and
\begin{equation}
    \bm{l}=\frac{1}{\sqrt{2}N}\left[ \partial_\tau+(N-v)\partial_\ell\right].
\end{equation}
The covariant components associated with these vectors are
\begin{equation}
    k_a=(-(N+v),-1,0,0)_a,
\end{equation}
\begin{equation}
    l_a=(-(N-v),1,0,0)_a,
\end{equation}
Hence, the transverse metric defined in Eq.~\eqref{eq:htrans} satisfies
\begin{equation}
    h_{\tau\tau}=h_{\tau\ell}=h_{\ell\ell}=0,
\end{equation}
which implies that $h_{ab}$ is reduced to the usual angular metric multiplied by $r^2$, namely
\begin{equation}
h_{ab} = \left[ \begin{array}{cc|cc} 0 &0 &0&0\\ 
0 &0 & 0 &0 \\ \hline
0&0&r(\tau,\ell)^2 & 0\\0&0&0&r(\tau,\ell)^2\sin^2\theta\end{array}\right].
\end{equation}
The null expansion scalars are then given by
\begin{equation}
    \theta^{(\bm{k})}=\frac{\sqrt{2}}{r}\left\{ \frac{\partial r}{\partial \tau}-(N+v)\frac{\partial r}{\partial \ell}\right\},
\end{equation}
and
\begin{equation}
    \theta^{(\bm{l})}=\frac{\sqrt{2}}{r}\left\{\frac{\partial r}{\partial \tau}+(N-v)\frac{\partial r}{\partial \ell}\right\}.
\end{equation}
Therefore, $\theta^{(\bm{s})}$ is given by
\begin{equation}
    \theta^{(\bm{s})}=\frac{\theta^{(\bm{l})}-\theta^{(\bm{k})}}{\sqrt{2}}=\frac{2N}{r}\frac{\partial r}{\partial\ell},
\end{equation}
while
\begin{equation}
    \theta^{(\bm{n})}=\frac{2\sqrt{2}}{r}\left(\frac{\partial r}{\partial \tau}-v\frac{\partial r}{\partial \ell}\right).
\end{equation}

Taking into account that $N$ must be non-zero, the condition in the definition of a UMOTS that $\left.\theta^{(\bm{s})}\right|_{\mathscr{S}}=0$ is equivalent in these coordinates to
\begin{equation}\label{eq:suhdef3}
    \left.\frac{\partial r}{\partial \ell}\right|_{\mathscr{S}}=0.
\end{equation}
If we impose the remaining conditions that $\left.\theta^{(\bm{k})}\right|_{\mathscr{S}}<0$ and $\left.\theta^{(\bm{l})}\right|_{\mathscr{S}}<0$ (which then imply $\left.\theta^{(\bm{n})}\right|_{\mathscr{S}}<0$) it follows that the following relation must also be satisfied:
\begin{equation}\label{eq:suhdef4}
    \left.\frac{\partial r}{\partial \tau}\right|_{\mathscr{S}}<0.
\end{equation}
Let us now compare this definition with the usual definition of universal horizons valid in static situations~\cite{Barausse:2011pu,Blas2011}. We should obtain this as a limiting case of our definitions above, valid for more general situations. The location of the universal horizon is determined by the condition that the gradients of $\tau$ and $r$ are anti-parallel, namely
\begin{equation}\label{eq:suhdef1}
    \left.\nabla_a\tau\right|_{\mathscr{S}}=-\chi^2\left.\nabla_a r\right|_{\mathscr{S}},
\end{equation}
where $\chi^2$ is a positive proportionality constant. This relation translates into
\begin{equation}\label{eq:suhdef2}
    \left.\frac{\partial r}{\partial l}\right|_{\mathscr{S}}=0,\qquad \left.\frac{\partial r}{\partial \tau}\right|_{\mathscr{S}}<0.
\end{equation}
These are the two relations obtained above, namely Eqs.~\eqref{eq:suhdef3} and \eqref{eq:suhdef4}. 

An additional aspect that we have introduced above is the outer/inner distinction of universally marginally trapped surfaces, which led us to the definitions of UMOTS and UMITS. Aside from Eq.~\eqref{eq:suhdef2} being satisfied, an UMOT is characterized by
\begin{equation}\label{eq:umotsmin}
\left.\frac{\partial^2 r}{\partial l^2}\right|_{\mathscr{S}}>0,    
\end{equation}
while an UMIT is characterized by
\begin{equation}\label{eq:umitsmax}
\left.\frac{\partial^2 r}{\partial l^2}\right|_{\mathscr{S}}<0.    
\end{equation}
These relations will turn out to be useful when discussing specific geometries.

\section[Implications of the kinematical part of Penrose's theorem for Lorentz-violating theories]{Implications of the kinematical part of Penrose's theorem for Lorentz-violating theories \label{sec:reg}}

The goal of this paper is classifying all possible geodesically complete spacetimes describing spherically symmetric black holes in Lorentz-violating theories. We will follow the discussion in Refs.~\cite{Carballo-Rubio2019a,Carballo-Rubio2019b}. The main contribution of \cite{Carballo-Rubio2019a} is the classification of the possible deformations of the spacetime geometry around the focusing point that guarantee the completeness of null geodesics. Different deformations achieving this goal lead to different families of geometries, discussed in more detail below.

Hence, the starting point of the discussion is the kinematical part of Penrose's theorem, which implies the incompleteness of null geodesics due to the impossibility of having a focusing point at a finite affine distance. Given that in Lorentz-violating theories the trajectories of physical particles do not necessarily correspond to the trajectories inside a given metric, the existence of focusing points may not imply the existence of physical singularities in these more generic situations in which one can define several distinct metrics.

\subsection{Einstein--aether frameworks}

Let us start considering the situation in which all propagating modes have finite signal velocities. The case in which the signal velocity is infinite for at least one of the propagating modes needs a separate treatment, which will be discussed later. 

In Einstein--aether frameworks, for each propagating mode there exists a metric~\eqref{eq:metmod}. As a consequence of the kinematical part of Penrose's theorem, we can conclude that none of these metrics can have focusing points. Indeed, if one of these metrics had a focusing point, then we could conclude the geodesic incompleteness of this metric, which signals the existence of a singular behavior for the corresponding propagating modes.

Let us consider the metric $\hat{g}_{ab}$ associated with the largest signal velocity. As discussed previously, OTHs of $\hat{g}_{ab}$ allow us to define the boundary of a black hole. Due to the finite values of signal velocities in these frameworks, it is not necessary to introduce UOTHs to define black holes. However, it is interesting to notice that known black hole solutions in Einstein--aether indeed have universal horizons \cite{Barausse:2011pu}. In these frameworks, UOTHs are inside the OTHs of the metric associated with the largest signal velocity, and therefore UOTHs do not define the boundary of black holes in these theories.

Taking this into account, we can conclude the following in relation to the classification presented in~\cite{Carballo-Rubio2019b}. If no UOTHs are present, the metric $\hat{g}_{ab}$ associated with the largest signal velocity must belong to one of the classes in~\cite{Carballo-Rubio2019b}. If UOTHs are present, the metric associated with the largest signal velocity must belong to one of the classes discussed in Sec.~\ref{sec:class} for Ho\v rava--like frameworks. 

A novel issue that appears due to the existence of different metrics is that it is not straightforward to conclude that the remaining metrics will belong to the same class as the metric associated with the largest signal velocity. However, we know that the expansion scalars associated with different metrics are related by Eq.~\eqref{eq:likimts}, and there could be additional relations of this sort that might be enough to show that different metrics must belong to the same classes.

\subsection{Ho\v{r}ava--like frameworks}

In the previous subsection we have discussed frameworks in which all signal velocities are finite. We now consider the case in which infinite signal velocities are allowed. For simplicity, we consider the situation in which a single signal velocity is infinite, while the remaining propagating modes have finite signal velocities.

For each of the the propagating modes with finite signal velocities, we can still define the metric in Eq.~\eqref{eq:metmod}. As in the previous section, we can consider the mode with the highest (finite) signal velocity $\hat{g}_{ab}$. For a black hole to be present in these frameworks, a UOTH must exist which, following our discussion in Sec.~\ref{sec:local}, implies the existence of a OTH for the metric $\hat{g}_{ab}$. That the propagating modes associated with $\hat{g}_{ab}$ do not experience any singular behavior implies the lack of focusing points and, therefore, that $\hat{g}_{ab}$ belongs to one of the classes in Refs.~\cite{Carballo-Rubio2019a,Carballo-Rubio2019b}, with the additional constraint of the existence of a UOTH within the trapped region. The consequences that follow from this additional ingredient are discussed in detail in Sec.~\ref{sec:class}.

On the other hand, we still need to analyze the trajectories of particles with infinite velocities and make sure that no singularities are found along these trajectories. It seems reasonable to expect that these additional conditions should translate into additional constraints on the fields at hand, for instance the preferred vector field $n^a$. To provide an answer to this issue we first need to determine the trajectories followed by these particles.

Particles with infinite velocities cannot follow geodesics of the spacetime metric. Hence, it is interesting to find the alternative equations that determine the motion of these particles. We have both the spacetime metric $g_{ab}$ and the irrotational aether $n_a \propto \nabla_a \tau$ to play with, so we should be able to build some dynamical equation using only those fields.
The infinite-velocity particles must be following spacelike curves, so you can use spacelike proper distance $ds = \sqrt{g_{ab} dx^a dx^b}$ to parameterize them and define a spacelike 4-velocity $V^a = dx^a/ds$.

One can certainly calculate the 4-acceleration $A^a = V^b \nabla_b V^a$, which generically is not zero. However, we need to determine an equation to be satisfied by this acceleration. The only other preferred vector we have is $n_a =  \nabla_a \tau/||\nabla \tau||$. 
Hence, the only plausible replacement for the geodesic equation is this
\begin{equation}
A^a = V^b \nabla_b V^a \propto n^a,
\end{equation}
which we can rewrite as
\begin{equation}
V^b \nabla_b V^a = \alpha \; n^a.
\end{equation}

Let us now calculate the proportionality factor $\alpha$. 
By the definition of infinite velocity particle we have $V^a \nabla_a \tau = 0$, i.e. they are supposed to be moving on constant chronon hypersurfaces. So $V^a n_a = 0$. Now differentiate this condition
\begin{equation}
0 = \nabla_b (V^a n_a) = (\nabla_b V^a) n_a + V^a \nabla_b n_c = 0.
\end{equation}
Contract with $V^b$:
\begin{equation}
V^b (\nabla_b V^a) n_a = - V^b V^a \nabla_b n_a.
\end{equation}
That is
\begin{equation}
A^a  n_a = - V^b V^a \nabla_b n_a.
\end{equation}
Thence by our assumption 
\begin{equation}
\alpha \; n^a n_a = - n_{(a;b)} V^a V^b.
\end{equation}
That is
\begin{equation}
\alpha  = n_{(a;b)} V^a V^b.
\end{equation}
And so our infinite-velocity particles have 4-acceleration
\begin{equation}
A^a = V^b \nabla_b V^a = \{ n_{(b;c)} V^b V^c \} \; n^a. 
\end{equation} 

But since the 4-velocities $V$ are perpendicular to $n$, and lie in the constant chronon hypersurface, this means we can replace $n_{(b;c)} $ by the extrinsic curvature $K_{bc}$ so that from the 4-dimensional point of view 
\begin{equation}
\label{E:4d}
A^a = V^b \nabla_b V^a =  \{ K_{bc} V^b V^c \} n^a
\end{equation}

That is
\begin{equation}
 V^b \partial_b V^a +  ^{(4)\!}\Gamma^a{}_{bc} V^b V^c =    \{ K_{bc} V^b V^c \} n^a.
\end{equation}
But for infinite velocity curves $V^a=(0;V^i)$, so looking at the spatial components
\begin{equation}
 V^j \partial_j V^i +  ^{(4)\!}\Gamma^i{}_{bc} V^b V^c =    \{ K_{bc} V^b V^c \} n^i.
\end{equation}
Since $V^0=0$ this further simplifies to
\begin{equation}
 V^j \partial_j V^i +  ^{(4)\!}\Gamma^i{}_{jk} V^j V^k =    \{ K_{jk} V^j V^k \} n^i.
\end{equation}
That is
\begin{equation}
 V^j \partial_j V^i +  (^{(4)\!}\Gamma^i{}_{jk} - n^i K_{jk} ) V^j V^k =  0.
\end{equation}
But this now has a nice 3-dimensional interpretation:
\begin{equation}
V^j \partial_j V^i  + ^{(3)\!}\Gamma^i{}_{jk} \;V^j V^k = 0.
\end{equation}
That is, the infinite velocity curves follow 3-dimensional geodesics of the intrinsic geometry.

For completeness, and as a consistency check, consider the time component of the 4-d equation (\ref{E:4d}). Using $n^0=1$ we see
\begin{equation}
0 + ^{(4)\!}\Gamma^0{}_{jk} V^j V^k = K_{jk} \; V^j V^k.
\end{equation}
Since this holds for arbitrary $V^i$ we see that (as expected) 
\begin{equation}
^{(4)\!}\Gamma^0{}_{jk}  = K_{jk}. 
\end{equation}

Hence, the following statements are equivalent: If infinite velocity curves are governed by ``4-acceleration is proportional to the gradient of  the chronon" then this is equivalent to ``infinite velocity curves follow  3-dimensional geodesics of the intrinsic geometry of the constant-chronon hypersurfaces''.
This is ultimately very natural since it implies that infinite velocity curves really only see the particular constant chronon hypersurface they are living on, and don't care about the rest of the spacetime.

\section{Classification of possible geometries \label{sec:class}}

We can now present the possible geometries that describe geodesically complete black holes in Lorentz-violating frameworks. As we have discussed above, any of the metrics associated with propagating modes with finite signal velocities must belong to one of the classes in~\cite{Carballo-Rubio2019b}. This is true for both Einstein--aether and Ho\v rava--like frameworks. In the latter frameworks, it is necessary that an UOTH exists (in Einstein-aether frameworks this is possible, but not necessary).

Hence, in this section we focus on discussing in detail the classes of geometries for which an UOTH exists. Let us pick a specific propagating mode and the associated metric $g^{(i)}_{ab}$. We can consider an ingoing null geodesic that intersects the UOTH at a given spacetime point with radius $r=r_\star$. In practice, this means that we are considering a foliation in slices of constant advanced null time $v$, although we can equally switch to the foliation in terms of constant-chronon slices if convenient for the discussion. The UOTH is a 3-surface foliated by UMOTS which implies, from the definition of the latter in Eq.~\eqref{eq:umtsdef}, that in the slice we are considering we have [Eq. \eqref{eq:qluhdef2}]
\begin{equation}
    \left.\theta^{\left(\bm{l}^{(i)}\right)}\right|_{\mathscr{S}}=\left.\theta^{\left(\bm{k}^{(i)}\right)}\right|_{\mathscr{S}}<0.
\end{equation}
As we will be following the ingoing null geodesic starting from the point $r=r_\star$ in the geodesic, it is useful to label it in terms of the specific value $\lambda=\lambda_\star$ of an affine parameter. At each point along the ingoing null geodesic, there will be outgoing null geodesics intersecting the former. As $\theta^{\left(\bm{s}^{(i)}\right)}$ becomes negative just behind the UMOTS in this slice, using Eq.~\eqref{eq:nsexplk} we can deduce that there always exists an open interval $(\lambda_\star,\lambda_{\star\star})$ in which the following equation is satisfied:
\begin{equation}\label{eq:uhimp}
    \theta^{\left(\bm{l}^{(i)}\right)}(r(\lambda))<\theta^{\left(\bm{k}^{(i)}\right)}(r(\lambda))<0,\qquad \lambda\in(\lambda_\star,\lambda_{\star\star}).
\end{equation}
In terms of $\theta^{\left(\bm{n}^{(i)}\right)}$ and $\theta^{\left(\bm{s}^{(i)}\right)}$, we have then
\begin{equation}\label{eq:uhimp2a}
    \theta^{\left(\bm{n}^{(i)}\right)}(r(\lambda))<0,\qquad \theta^{\left(\bm{s}^{(i)}\right)}(r(\lambda))<0,\qquad \lambda\in(\lambda_\star,\lambda_{\star\star}),
\end{equation}
which taking into account Eq.~\eqref{eq:nsexplk} implies
\begin{equation}\label{eq:uhimp2}
    \theta^{(\bm{n})}(r(\lambda))<0,\qquad \theta^{(\bm{s})}(r(\lambda))<0,\qquad \lambda\in(\lambda_\star,\lambda_{\star\star}).
\end{equation}
On the other hand, the possible behavior of the null expansions were determined for the different families of geometries discussed in~\cite{Carballo-Rubio2019a,Carballo-Rubio2019b}. The discussion in~\cite{Carballo-Rubio2019a,Carballo-Rubio2019b} was general enough so that it can be applied to the present discussion. The only difference is that now we will restrict our attention to the geometries with UOTHs, which in practical terms is encapsulated in Eq.~\eqref{eq:uhimp} being satisfied. Hence, we just need to proceed case by case following the classification of reference \cite{Carballo-Rubio2019a,Carballo-Rubio2019b}, and discussing the implications that stem from Eq.~\eqref{eq:uhimp}.

As discussed in reference \cite{Carballo-Rubio2019b}, the most generic geometry can be classified in terms of the three parameters $\left(\lambda_{\rm defocus}, R_{\rm defocus},\theta^{(\bm{k})}(r\left(\lambda_{\rm defocus}\right))\right)$, where $\lambda_{\rm defocus}$ indicates the value of the affine parameter for which the expansion relative to the outgoing null geodesic changes sign, $R_{\rm defocus}=r\left(\lambda_{\rm defocus}\right)$ indicates the areal radius at which this defocusing occurs, and $\theta^{(\bm{k})}(r\left(\lambda_{\rm defocus}\right))$ indicates the sign of the expansion relative to the ingoing null geodesic at the defocusing point. Classifying the allowed geometries in terms of these three parameters was the most convenient choice due to having a single metric to deal with. However, for the theories analyzed in this paper we have several metrics $g^{(i)}_{ab}$, and each of them may belong to a different class. 

On the other hand, we know from Eq.~\eqref{eq:nsexplk} that the expansions $\theta^{\left(\bm{s}^{(i)}\right)}$ and $\theta^{\left(\bm{n}^{(i)}\right)}$ are the same for all metrics, up to constant rescalings for the latter. Hence, if we were able to use these quantities to label the different families, we would know that all metrics $g^{(i)}_{ab}$ must belong to the same class. We will see below that, in fact, all possible regular geometries are characterized by the zeroes of the functions $\theta^{\left(\bm{n}^{(i)}\right)}$ or $\theta^{\left(\bm{s}^{(i)}\right)}$, which thus allows us to implement this idea.

\subsection{\textbf{Evanescent horizons}}

This class of geometries is characterized by the existence of a defocusing point, in which $\theta^{\left(\bm{l}^{(i)}\right)}$ vanishes, at a finite affine distance $\lambda_{\rm defocus}=\lambda_0$. On the other hand, $\theta^{\left(\bm{k}^{(i)}\right)}$ remains negative. This implies that
\begin{equation}\label{eq:thetas}
   \left. \theta^{(\bm{s})}(r(\lambda))\right|_{\lambda=\lambda_0}>0.
\end{equation}
Hence, recalling Eq.~\eqref{eq:uhimp2} we can conclude that this class is characterized by a change of sign of $\theta^{(\bm{s})}$ which, therefore, must vanish at least once before $\lambda=\lambda_0$. In other words, $\theta^{(\bm{s})}$ must display an even number of zeroes (with one of them being the one associated to the UMOTS that we have assumed to exist).

\begin{figure}[!ht]%
\begin{center}
\vbox{\includegraphics[width=0.7\textwidth]{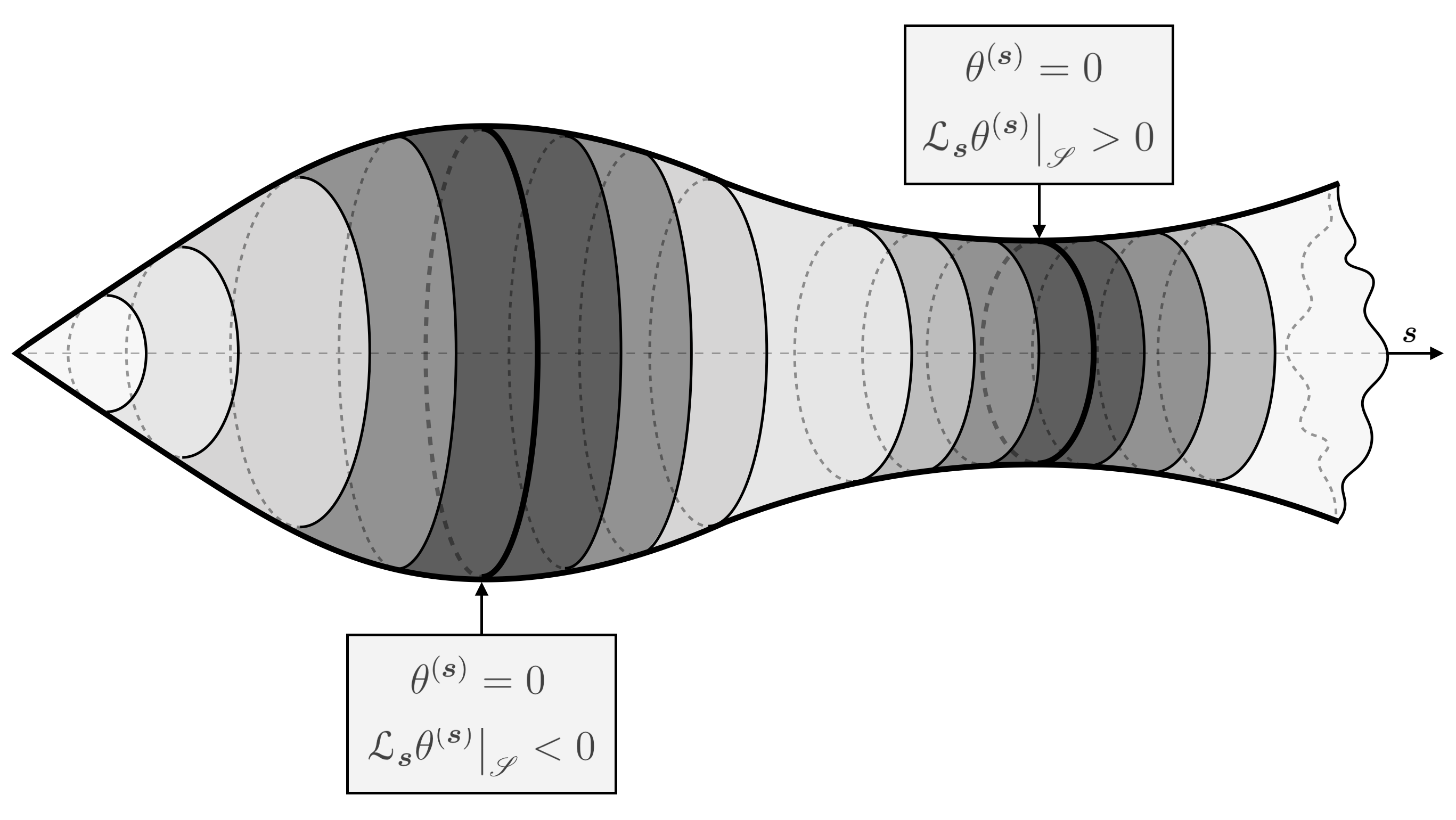}}
\bigskip%
\caption{Representation of the structure of a UOTH/UITH pair as seen within a constant-chronon hypersurface, with the circular sections representing the areal radius $r$ and the outward direction indicated by $\bm{s}$. The UMOTS marks a local minimum in the areal radius, while the UMITS marks a local maximum. Around each of these local extrema of the areal radius, one can see the nested structure discussed in more detail in Fig.~\ref{fig:nested}. There are regions between the UMOTS and the UMITS that do not contain trapped surfaces for the metric $g^{(i)}_{ab}$ for high enough values of $c_{(i)}/c$. However, the areal radius is bounded from above for propagating modes in these regions while the pair UOTH/UITH exists, so that these regions are still trapped in this looser sense.}
\label{fig:chrononminmax}
\end{center}
\end{figure}%

The simplest geometries in this class are those in which $\theta^{(\bm{s})}$ vanishes just once before reaching $\lambda=\lambda_0$. This implies the existence of at least one other position on the slice (aside from the UMOTS) at which $\theta^{(\bm{s})}$ vanishes so that, overall, $\theta^{(\bm{s})}$ vanishes twice. 
In this case, this additional position marks the location of an UMITS. To see this, let us recall that an UMOTS satisfies Eq.~\eqref{eq:umotsmin} (namely, $\left.\partial^2 r / \partial l^2\right|_{\mathscr{S}}>0$). If we consider the constant-chronon slices, then an UMOTS is coincident with a local minimum of the areal radius. Hence, after crossing the UMOTS the areal radius increases. Due to continuity reasons, and the fact that the constant-chronon hypersurfaces have topology $\mathbb{R}^3$, there must exist a local maximum of the areal radius. But this needs to be coincident with the other position in which $\theta^{(\bm{s})}$ vanishes, which taking into account Eq.~\eqref{eq:umitsmax} (namely, $\left.\partial^2 r / \partial l^2\right|_{\mathscr{S}}<0$) then marks the location of an UMITS (see Fig.~\ref{fig:chrononminmax}).

\begin{figure}[!ht]%
\begin{center}
\vbox{\includegraphics[width=0.3\textwidth]{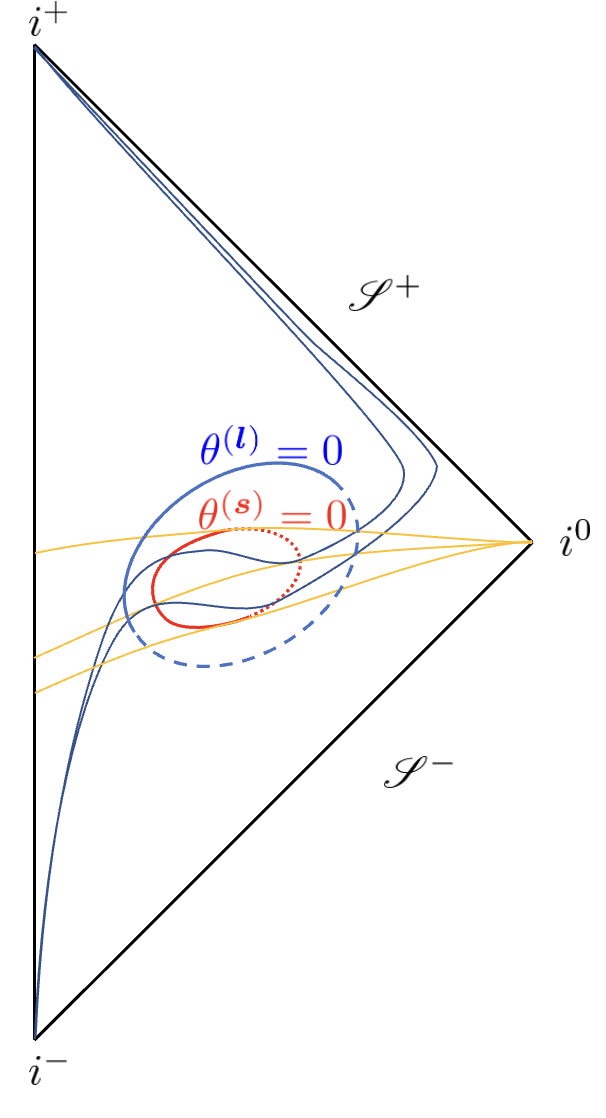}}
\bigskip%
\caption{Penrose diagram for an evanescent horizon black hole with a single UOTH/UITH pair. The dotted red line denotes the UOTH, while the solid red line denotes the UITH. For any metric $g^{(i)}_{ab}$, there must also exist a OTH  indicated by the blue dashed line and a ITH indicated by the blue solid line. All the metrics $g^{(i)}_{ab}$ belong to the same class and the position of the UOTH and UITH are the same, while the positions of the OTH and ITH are shifted accordingly to the value of $c_{(i)}$, resulting in a nested structure such as the one in Fig.~\ref{fig:nested} but for both OTHs and ITHs. The combination of UOTH and UITH form a closed structure, as it happens with outer and inner trapping horizons in regular black holes~\cite{Frolov:2014jva}. This diagram also provides a complementary illustration, from that in Fig.~\ref{fig:chrononminmax}, that if we follow constant-chronon hypersurfaces (yellow lines)  in the outgoing direction, once these hypersurfaces cross the UITH their areal radius must decrease until reaching the UOTH.
}
\label{fig:evanescent}%
\end{center}
\end{figure}%

\begin{figure}[!ht]%
	\centering
		\includegraphics[scale=0.5]{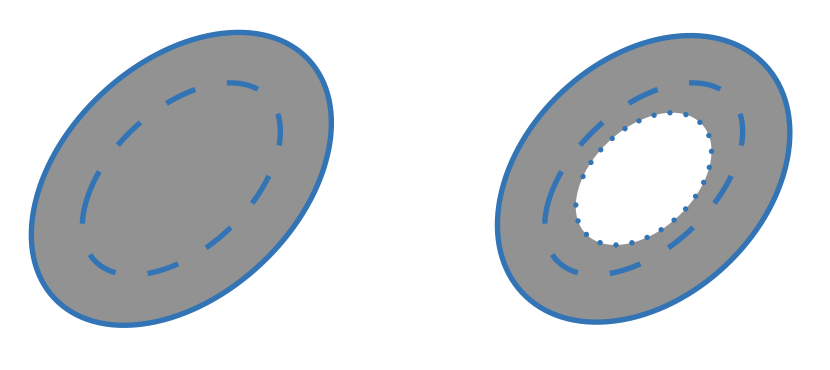}
	\protect\caption{Schematic representation of trapped regions as seen by propagating modes with different signal velocities. The solid lines represent regions in which $\theta^{\left(\bm{l}^{(i)}\right)}=0$, dashed lines for $\theta^{(\bm{s})}=0$, and dotted lines for $\theta^{\left(\bm{k}^{(i)}\right)}=0$. For low enough signal velocities (left figure) all the region bounded by the UOTH/UITH pair (and, by extension, the OTH/ITH pair associated wich each velocity) is a trapped region. On the other hand, as depicted in the figure on the right, for high enough signal velocities there exists a region (marked in white) that is not trapped inside the region bounded by the UOTH/UITH pair. This internal white region is bounded by an hypersurface in which $\theta^{\left(\bm{k}^{(i)}\right)}$ vanishes. 
}
\label{fig:trap_reg}%

\end{figure}

More generally, $\theta^{(\bm{s})}$ can vanish an even number of times, with the corresponding geometries being characterized by UOTHs/UITHs pairs, similarly to the situation described in~\cite{Carballo-Rubio2019b} for OTHs/ITHs. The simplest example consists of a geometry with an inner and outer horizon that merge in finite time. Geometries of this kind describe a regular black hole that disappears in finite time. To our knowledge, it is the first time that it has been pointed out that the description of non-singular black holes in Lorentz-violating theories must involve UITH aside from a UOTH. The Penrose diagram of an evanescent non-singular black hole with a UITH is presented in Fig. \ref{fig:evanescent}.
Fig.~\ref{fig:trap_reg} illustrates the different behavior of the trapped regions for high and low velocity energy modes.  Modes that propagate at sufficiently high velocity are characterized by a region where the expansion relative to the ingoing direction vanishes and changes sign. Such region can be absent for modes propagating at low velocities.

Let us stress that the crucial characteristic of this class is that horizons come in pairs. This applies both to horizons foliated by either future or past marginally trapped surfaces. This implies that we can construct more complicated geometries in this class with additional (future or past) trapped surfaces, with the only constraint that these regions are bounded and do not mutually intersect. In fact, the situation is completely parallel to the discussion in~\cite{Carballo-Rubio2019b}, but replacing MOTS/MITS with UMOTS/UMITS. These geometries would describe regular black holes that transform into white holes an arbitrary number of times (that depends on the number of trapped regions in the geometry). Let us recall that, in the present discussion, the outer boundary of white holes are 3-surfaces foliated by past universally marginally trapped surfaces as defined in Eq.~\eqref{eq:umtsdef}.

Notice that a given metric belongs to this class if and only if $\theta^{\left(\bm{s}^{(i)}\right)}$ vanishes an even number of times in a given slice, and that $\theta^{\left(\bm{s}^{(i)}\right)}=\theta^{(\bm{s})}$ is an invariant quantity under changes of $c_{(i)}$. Hence, it follows straightforwardly that, if a single specific metric $g^{(i)}_{ab}$ is shown to belong to this class, all the remaining metrics must belong to the same class.

\subsection{\textbf{One-way hidden wormholes}}

\begin{figure}[!ht]%
\begin{center}
\vbox{\includegraphics[width=0.5\textwidth]{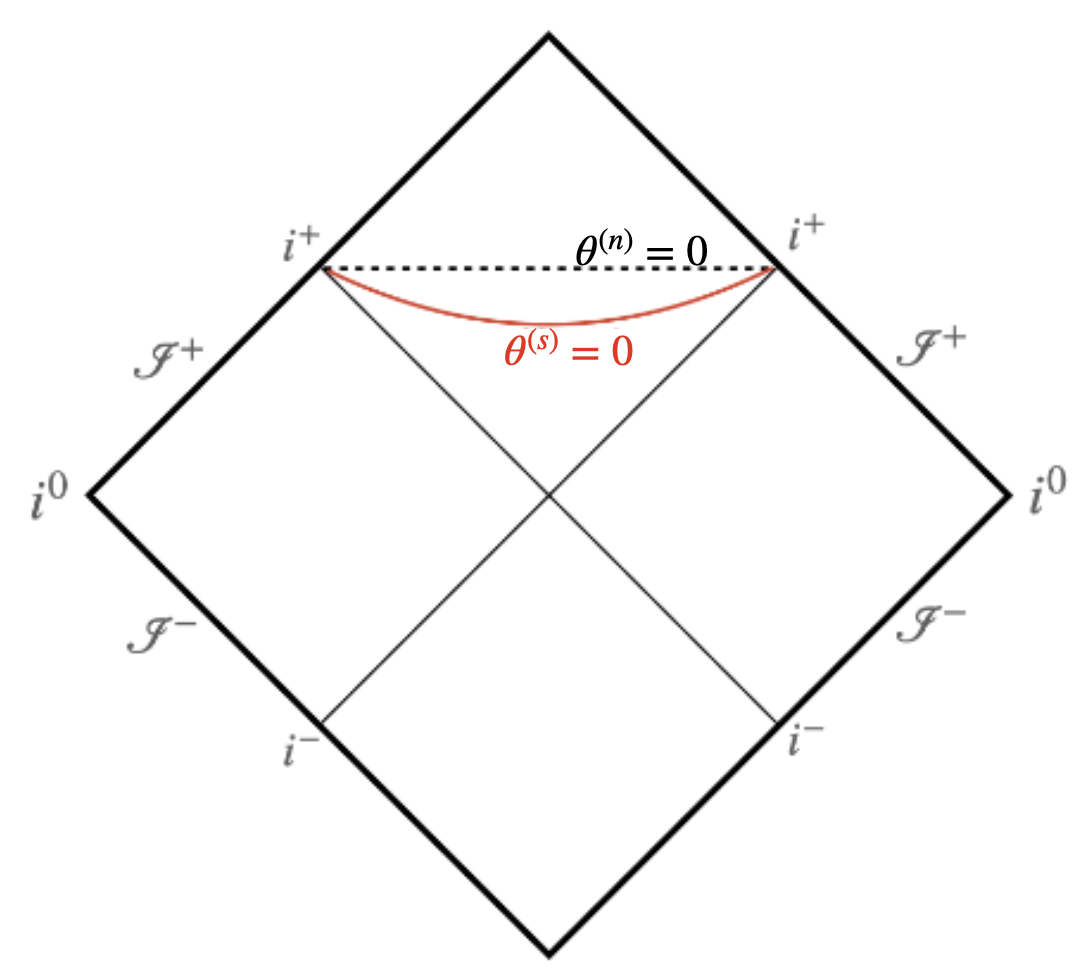}}
\bigskip%
\caption{Penrose diagram for an asymptotic hidden wormhole with universal horizon. The dotted line represents the wormhole throat at which $\theta^{\bm{(n)}}=0$. }
\label{fig:as_worm}%
\end{center}
\end{figure}

As in the previous case, also in this class of geometries there is a defocusing point for a finite value of the affine parameter at which $\theta^{\left(\bm{l}^{(i)}\right)}$ vanishes. 
The difference with respect to the evanescent horizons class is that $\theta^{\left(\bm{k}^{(i)}\right)}$ changes sign so that  $\left.\theta^{\left(\bm{k}^{(i)}\right)}\right|_{\lambda=\lambda_0}\geq0$. We know from our discussion in~\cite{Carballo-Rubio2019a} that geometries in this class are characterized by a (generically dynamical) wormhole throat which corresponds to a minimum radius hypersurface (see Fig.~\ref{fig:as_worm}). 

The fact that $\theta^{\left(\bm{k}^{(i)}\right)}$ changes sign implies that
\begin{equation}\label{eq:thetas_hw}
   \left. \theta^{(\bm{s})}(r(\lambda))\right|_{\lambda=\lambda_0}\leq 0.
\end{equation}
As a consequence, it is not necessary that the expansion $\theta^{(\bm{s})}$ changes sign for the geometries in this family. On the other hand, it follows that
\begin{equation}\label{eq:owhwn}
   \left. \theta^{(\bm{n})}(r(\lambda))\right|_{\lambda=\lambda_0}\geq 0.
\end{equation}
Hence, recalling Eq.~\eqref{eq:uhimp2} we can conclude that this class is characterized by a change of sign of $\theta^{(\bm{n})}$ which, therefore, must vanish at least once before $\lambda=\lambda_0$. Hence, in this case $\theta^{(\bm{n})}$ must display an odd numbers of zeroes in the interval that goes from the UMOTS to the defocusing point.

\begin{figure}[!ht]%
\begin{center}
\vbox{\includegraphics[width=0.5\textwidth]{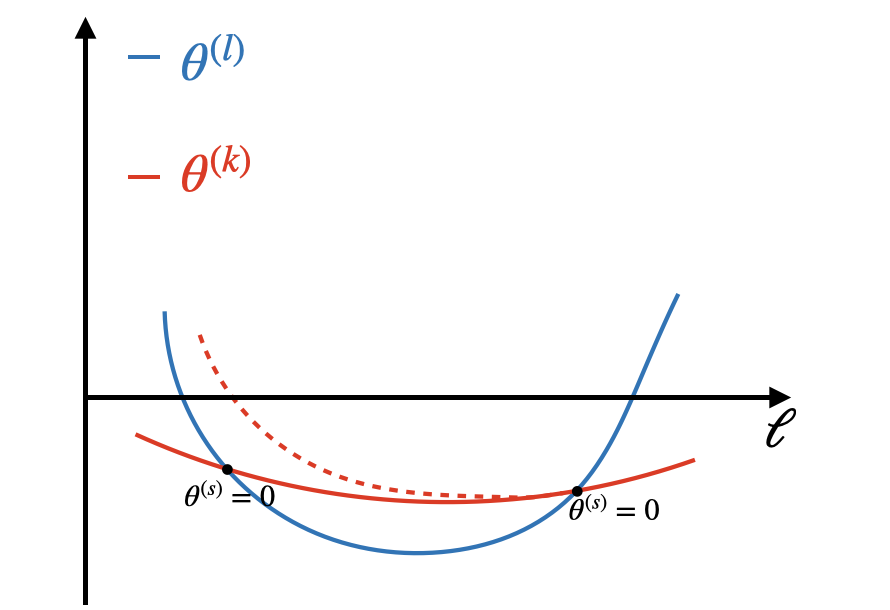}}
\bigskip%
\caption{Qualitative behavior of the expansions  $\theta^{\left(\bm{l}^{(i)}\right)}$ and $\theta^{\left(\bm{k}^{(i)}\right)}$ for a fixed value of time as a function of the proper distance $\ell$ from the center of spacetime. In the case of the evanescent horizon class (solid line) and the one-way hidden wormhole case (dotted line). $\theta^{\left(\bm{l}^{(i)}\right)}$ becomes smaller than $\theta^{\left(\bm{k}^{(i)}\right)}$ inside the universal horizon for both classes of spacetimes. In the evanescent horizon class, the expansions switch order again closer to the center with $\theta^{\left(\bm{l}^{(i)}\right)}$ eventually changing sign. In the one-way hidden wormhole class, $\theta^{\left(\bm{k}^{(i)}\right)}$ remains greater the $\theta^{\left(\bm{l}^{(i)}\right)}$ and it eventually changes sign.}
\label{fig:expansions}%
\end{center}
\end{figure}

The simplest geometries in this class are those in which $\theta^{(\bm{n})}$ vanishes just once before reaching $\lambda=\lambda_0$. A qualitative picture of the behavior of the expansions $\theta^{\left(\bm{l}^{(i)}\right)}$ and $\theta^{\left(\bm{k}^{(i)}\right)}$ is depicted in Fig.~\ref{fig:expansions}. The limiting case in which $\left.\theta^{\left(\bm{k}^{(i)}\right)}\right|_{\lambda=\lambda_0}=0$ contains a static wormhole throat, which can be alternatively described as $\left.\theta^{(\bm{n})}\right|_{\lambda=\lambda_0}=\left.\theta^{(\bm{s})}\right|_{\lambda=\lambda_0}=0$. This situation is precisely the one shown in Fig.~\ref{fig:as_worm}.

As with the class above, the fact that geometries in this family are completely characterized by the points in which $\theta^{(\bm{n})}$ vanishes implies, together with Eq.~\eqref{eq:nsexplk}, that if a single specific metric $g^{(i)}_{ab}$ is shown to belong to this class, all the remaining metrics must belong to the same class.

\subsection{\textbf{Pushing the defocusing point to infinite affine distance}}

In the single-metric setting, two extra classes of geometries can be obtained by pushing the defocusing point to infinite affine distance along outgoing null geodesics~\cite{Carballo-Rubio2019a}. Whether these situations result in new classes for the multi-metric setting considered here must be analyzed independently.

\begin{figure}[!ht]%
\begin{center}
\vbox{\includegraphics[width=0.35\textwidth]{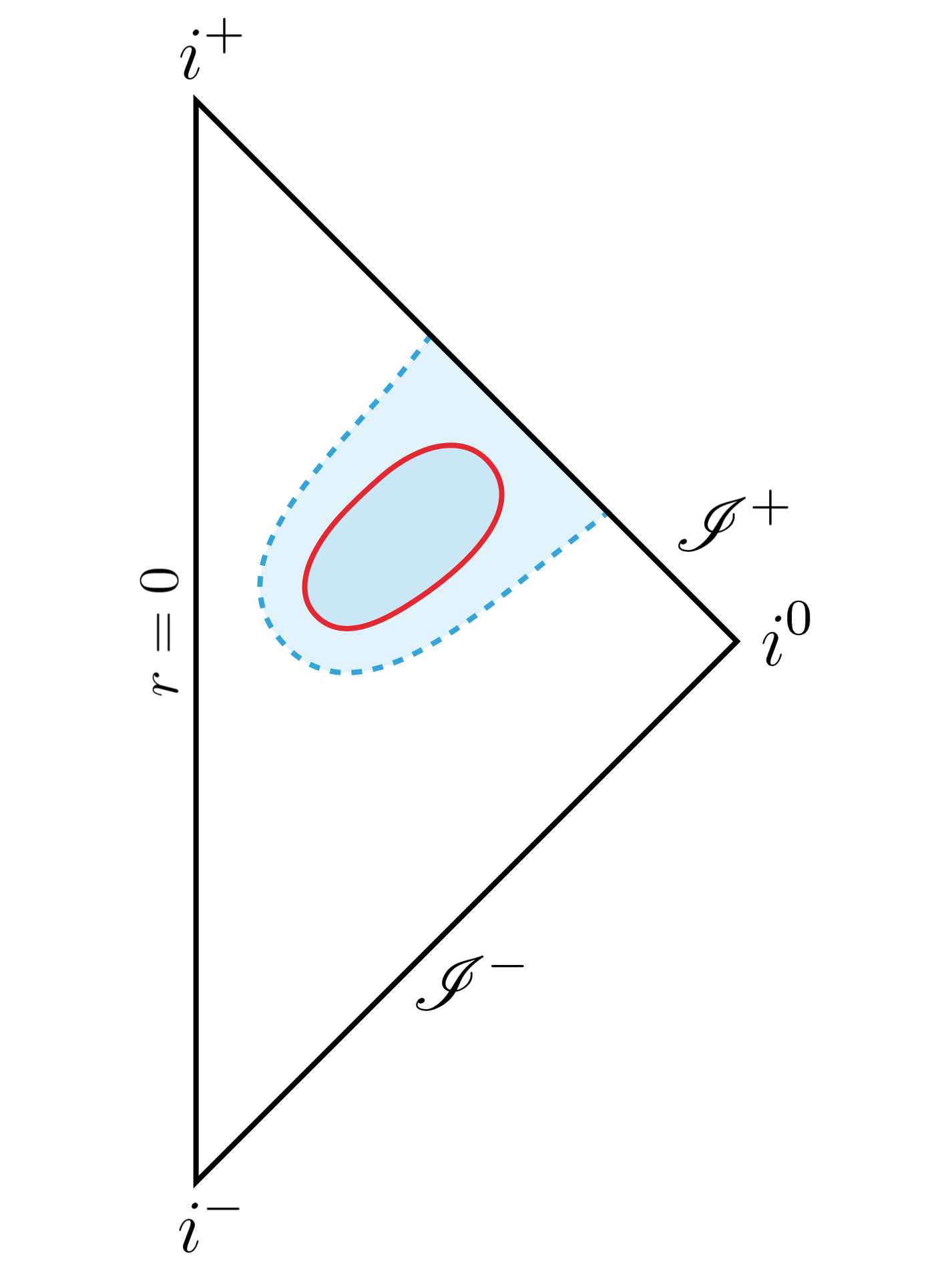}}
\bigskip%
\caption{Penrose diagram for a geometry in the evanescent horizons class that is characterized by the everlasting quality of trapping horizons for modes with signal velocities $c_{(i)}$ and lower. We have used solid lines for UMOTS/UMITS and dashed lines for the MOTS/MITS of $g^{(i)}_{ab}$.}
\label{fig:everlasting}%
\end{center}
\end{figure}

Let us first consider the $\lambda_0\rightarrow\infty$ limit of the evanescent horizons class described above, so that the defocusing point for $\theta^{\left(\bm{l}^{(i)}\right)}$ is pushed to infinite affine distance along the outgoing null geodesics of $g^{(i)}_{ab}$. This implies that
\begin{equation}
\lim_{\lambda_0\rightarrow\infty}\left. \theta^{(\bm{s})}(r(\lambda))\right|_{\lambda=\lambda_0}>0.
\end{equation}
Hence, $\theta^{\left(\bm{s}\right)}$ must vanish an even number of times, which implies that this class is also characterized by UOTH/UITH pairs. This seems to be the same situation as in the evanescent horizons case, and in fact it is equivalent from the perspective of UMOTS/UMITS, which merge in finite time. The difference here is that some of the low-energy modes remain trapped indefinitely (see Fig. \ref{fig:everlasting}). These geometries are thus a subset of the evanescent horizons class.

On the other hand, we can also consider the $\lambda_0\rightarrow\infty$ limit of the one-way hidden wormhole class obtaining a geometry associated to an asymptotic one-way hidden wormhole. Recalling Eq.~\eqref{eq:owhwn}, it follows that these geometries are characterized by
\begin{equation}
\lim_{\lambda_0\rightarrow\infty}\left. \theta^{(\bm{n})}(r(\lambda))\right|_{\lambda=\lambda_0}\geq0.
\end{equation}
If the limit is strictly positive these geometries are a subset of the one-way hidden wormholes class, as the zeroes of $\theta^{(\bm{n})}$ must display the same behavior. On other hand, if the equality holds we have a new class of geometries in which the wormhole throat is only reached asymptotically.

\subsection{\textbf{Singular geometries}}
As shown in Refs~\cite{Carballo-Rubio2019a,Carballo-Rubio2019b}, all other possible classes of geometries either lead to curvature singularities or to geodesic incompleteness. Instead of replacing the focusing point with a defocusing point, these singular classes push the focusing point to infinite affine distance. Therefore, the most generic non-singular geometry with trapping horizon is given by a combination of the geometries described in the previous sections.

It is illustrative to consider specific examples in this class. A particular way to construct a geometry in this class is taking one point in the UMOTS/UMITS and push it to infinite affine distance, which would be the equivalent of the behavior of MOTS/MITS in the everlasting horizons class discussed in~\cite{Carballo-Rubio2019a}. Hence, we would have
\begin{equation}
\lim_{\lambda_0\rightarrow\infty}\left. \theta^{(\bm{s})}(r(\lambda))\right|_{\lambda=\lambda_0}=0.
\end{equation}
This relation implies that either both $\theta^{\left(\bm{l}^{(i)}\right)}$ and $\theta^{\left(\bm{k}^{(i)}\right)}$ becomes non-negative, which is already covered in the one-way hidden wormholes or asymptotic hidden wormhole cases, or that $\theta^{\left(\bm{l}^{(i)}\right)}$ remains negative. The latter case can describe precisely a geometry in which the focusing point is pushed to infinite distance which, as shown in~\cite{Carballo-Rubio2019a}, results in a singularity at the center of spherical symmetry.

\section{Discussion and conclusions}\label{Concl}

In this paper we have analyzed the structure of geodesically complete black hole spacetimes in the context of Lorentz-violating gravity. We have classified all possible geometries in frameworks with modified dispersion relations, being our main conclusions the following:
\begin{itemize}
    \item Einstein-aether frameworks: These frameworks are characterized by finite signal velocities $c_{(i)}$ and a metric field $g^{(i)}_{ab}$ that can be defined for each propagating mode. We have concluded that each of these metrics must belong to one of the classes discussed in our previous papers (evanescent horizons, one-way hidden wormholes, everlasting horizons and asymptotic hidden wormholes)~\cite{Carballo-Rubio2019a,Carballo-Rubio2019b} with no amendments.
    \item Ho\v rava-like frameworks: These frameworks are characterized by the existence of at least one mode with infinite signal velocity. For each propagating mode with finite signal velocity $c_{(i)}$, it is still possible to construct a metric field $g^{(i)}_{ab}$. We have shown that the classes to which each of these metrics must belong are in one-to-one correspondence with three of the classes discussed in our previous papers (evanescent horizons, one-way hidden wormholes, and asymptotic hidden wormhole)~\cite{Carballo-Rubio2019a,Carballo-Rubio2019b}, but using the notion of universal trapping horizons instead of trapping horizons. From this perspective, infinite signal velocities simplify the classification of possible geometries, removing the limiting cases that were described by the everlasting horizons class.
\end{itemize}

These results illustrate the robustness and usefulness of our classification of geodesically complete black hole spacetimes. Some of the classes we have discussed are preserved when going from a single-metric framework to a multi-metric framework, and even if taking the limit in which some of the signal velocities are infinite. This is associated with the fact that, if we gradually modify the value of a given signal velocity so that it becomes infinite, the standard definition of trapping horizons is deformed smoothly the definition of universal trapping horizons we have discussed here. However, this limiting procedure works only if trapping horizons are compact. We have shown that it does not work for geometries that describe trapping horizons that never disappear, as it is not possible to find regular geometries in which universally trapping horizons never disappear.

It may seem surprising that frameworks with infinite signal velocities are as rich as situations with finite signal velocities, in the sense that it is possible to define for instance evaporating regular black holes, regular black holes bouncing into regular white holes, and hidden wormholes, but this is ultimately associated with the fact that it is possible to define a notion that replaces trapping horizons in these frameworks.

We believe that our results will be useful for the exploration of solutions describing geodesically complete black hole solutions in Ho\v rava-like frameworks. Theories of this sort are promising candidates to contain non-singular solutions, which if existing must belong to one of the classes in our catalogue. Given that our classification is purely kinematical, a clear question is whether specific dynamical behaviors can select specific classes as the preferred ones, and up to which extent this may be theory-dependent.

\begin{acknowledgments}
\noindent
FDF would like to thank Shinji Mukohyama for stimulating conversations.
FDF acknowledges financial support by Japan Society for the Promotion of
Science Grants-in-Aid for Scientific Research No.~17H06359. 
SL acknowledges funding from the Italian Ministry of Education and  Scientific Research (MIUR)  under the grant  PRIN MIUR 2017-MB8AEZ. MV was supported by the Marsden Fund, via a grant administered by the Royal Society of New Zealand.
\end{acknowledgments}

\bibliography{refs}

\end{document}